\documentclass[a4paper,12pt]{article}
\usepackage[english]{babel}
\usepackage{amssymb,amsmath,amsfonts,amsthm,epsf}
\usepackage{hyphenat}
\usepackage{latexsym}
\usepackage{graphicx}
\usepackage{color}

\usepackage{hyperref}

\newcommand{\be}{\begin{equation}}
\newcommand{\ee}{\end{equation}}
\newcommand{\bd}{\begin{definition}}
\newcommand{\ed}{\end{definition}}
\newcommand{\bt}{\begin{theorem}}
\newcommand{\et}{\end{theorem}}
\newcommand{\bp}{\begin{proof}}
\newcommand{\ep}{\end{proof}}
\newcommand{\bea}{\begin{eqnarray}}
\newcommand{\eea}{\end{eqnarray}}
\newcommand{\ba}{\begin{array}}
\newcommand{\ea}{\end{array}}

\def\siml{{\ \lower-1.2pt\vbox{\hbox{\rlap{$<$}\lower6pt\vbox{\hbox{$\sim$}}}}\ }} 
\def\bfnabla{\mbox{\boldmath $\nabla$}}

\def\bfsigma{\mbox{\boldmath $\sigma$}}

\def\als{\alpha_{s}}
\def\al{\alpha}
\def\lQ{\Lambda_{\rm QCD}}

\newcommand{\nn}{\nonumber}

\def\bfnabla{\mbox{\boldmath $\nabla$}}

\newcommand{\eq}[1]{Eq.~(\ref{#1})}

\def\dsl{\,\raise.15ex\hbox{/}\mkern-13.5mu D}

\newcommand{\MS}{\overline{\rm MS}}

\numberwithin{equation}{section}

\begin{document}

\begin{titlepage}
\begin{flushright}
\end{flushright}

\vspace{1cm}
\begin{center}
\begin{Large}
{\bf The Lamb shift in muonic hydrogen and the proton radius from effective field theories}\\[2cm] 
\end{Large} 
{\large Clara Peset and Antonio Pineda}\\
\vspace{0.5cm}
{\it  Grup de F\'\i sica Te\`orica, Dept. F\'\i sica and IFAE, Universitat Aut\`onoma de Barcelona,\\ 
E-08193 Bellaterra (Barcelona), Spain}\\
\today
\end{center}

\vspace{1cm}

\begin{abstract}
We comprehensively analyse the theoretical prediction for the Lamb 
shift in muonic hydrogen, and the associated determination of the proton radius.  We use effective field theories. This allows us to relate the 
proton radius with well-defined objects in quantum field theory, eliminating unnecessary model dependence. 
The use of effective field theories also helps us to organize the computation so that we can clearly state the parametric 
accuracy of the result. 
In this paper we review all (and check several of) the contributions to the energy shift of order $\al^5$, as well as those that scales like $\al^6\times$logarithms in the context of non-relativistic effective field theories of QED. 
\vspace{5mm} 

\end{abstract}

\end{titlepage}

\vfill
\setcounter{footnote}{0} 
\vspace{1cm}
\newpage

\tableofcontents

\vfill
\newpage

\section{Introduction}

The measurement \cite{Pohl:2010zza,Antognini:1900ns} of the Lamb shift in muonic hydrogen, 
\begin{eqnarray}
\label{DeltaEexp}
&&E(2P_{3/2})-E(2S_{1/2})\equiv \Delta E_L^{\rm exp}=202.3706(23)\,\mathrm{meV}
\end{eqnarray}
and the associated determination of the root mean square electric radius of the proton:
$r_p \equiv \sqrt{\langle r_p^2 \rangle}= 0.84087(39)$ fm has led to a lot of controversy. The reason is that this number is 7.1$\sigma$ away from the CODATA value, $r_p= 0.8775(51)$ fm \cite{Mohr:2012tt}. This last number is an average of determinations coming from 
hydrogen spectroscopy and electron-proton scattering\footnote{The latter though has been 
challenged in Refs.~\cite{Lorenz:2014vha,Lorenz:2014yda}, and its exclusion would certainly diminish this tension.}.  
In order to asses the significance of the discrepancy, it is of fundamental importance to perform the computation of 
the Lamb shift in muonic hydrogen (in particular of the errors)
in a model independent way. This was done in Ref.~\cite{Peset:2014yha}. 
In that Letter we revisited the theoretical derivation of the Lamb shift using effective field theories (EFTs) and obtained the following expression 
\bea
\label{El1}
&&
\Delta E_L^{\rm th}=
\left[
206.070(13)-
5.2270(7)
\frac{ r_p^2}{\mathrm{fm^2}}
\right]
\,\mathrm{meV}
\,.
\eea
Using this result and Eq.~(\ref{DeltaEexp}) we then obtained  
\begin{equation}
\label{rpfinal}
r_p=0.8413(15) \, \mathrm{fm},
\end{equation}
which is at 6.8$\sigma$ variance with respect to the CODATA value. Therefore, 
the proton radius puzzle survived our model independent analysis. The good point now is that 
the EFT analysis allows us to have a parametric control of the 
uncertainties, which are of the order of uncomputed terms of 
${\mathcal O}(m_{\mu}\alpha^5\frac{m_{\mu}^3}{m_{\rho}^3},m_{\mu}\alpha^6)$. This parametric control of the 
uncertainties allowed us to obtain a model independent estimate of the error, which is dominated by hadronic effects. 

EFTs help organizing the computation by providing with power counting rules that asses the importance of the different contributions. This 
is specially so for the muonic hydrogen, as its
dynamics is characterized by several scales:
\be
\nn
m_{p} \sim m_{\rho},
\quad
m_{\mu} \sim m_{\pi} \sim m_r\equiv\frac{m_{\mu}m_p}{m_p+m_{\mu}},
\quad
m_r\al \sim m_e.
\ee
By considering ratios between them, the main expansion parameters are obtained:
\bea
&&
\label{ratio1}
\frac{m_{\pi}}{m_p} \sim \frac{m_{\mu}}{m_p} \approx \frac{1}{9}
\,,
\;
\frac{m_e}{m_r} \sim \frac{m_r\al}{m_r}\sim \frac{m_r\al^2}{m_r\al}\sim \al \approx \frac{1}{137}
\,.
\eea

For our evaluation we used potential non-relativistic QED (pNRQED) \cite{Pineda:1997bj}. Particularly relevant for us is Ref. 
\cite{Pineda:2004mx}, which contains detailed information on the application of pNRQED to the muonic hydrogen. 
Since pNRQED describes degrees of freedom with $E \sim m_{\mu}\al^2$, any other degree of freedom with larger energy is integrated out. This implies treating the proton and muon in a non-relativistic fashion
and integrating out pions (and Delta particles). This is the step of going from Heavy Baryon Effective Theory
(HBET) \cite{Jenkins:1990jv} to Non-Relativistic QED (NRQED) \cite{Caswell:1985ui}.  By integrating out the scale $m_{\mu}\al$, pNRQED is obtained 
and the potentials appear. Schematically the path followed is the following ($\Delta\equiv m_\Delta-m_p$): 
$$
{\rm HBChPT} \; \stackrel{(m_{\pi/\mu},\Delta)}{\Longrightarrow} {\rm NRQED} \; \stackrel{(m_{\mu}\al)}{\Longrightarrow} \; {\rm pNRQED}\,.
$$
A detailed explanation of the matching computation between HBET and NRQED was given in Ref.~\cite{Peset:2014jxa}. This corresponds to 
the hadronic part of the computation presented in Ref.~\cite{Peset:2014yha}. It is one of the main motivations of
this paper to give the details of QED-related part of the analysis in Ref.~\cite{Peset:2014yha}. This means to analyse the potentials that contribute to the given order, as well as to actually compute the associated energy 
shifts associated to the potentials and the ultrasoft photons. We have made some effort to present the result 
assuming an arbitrary charge for the muon and proton, so that the results can be of use in a more general situation, in particular for muonic atoms. This is so because the expressions of the potentials would be equal for light muonic atoms after appropriately changing the NRQED Wilson coefficients produced by the hadronic scales. Therefore, we will present some results in terms of $Z_{\mu}(=1)$,  $Z_p(=1)$ and $Z \equiv Z_{\mu}Z_p(=1)$. 
We also expect that the analysis presented in this paper 
will set the basis for higher order computations using EFTs. 

\section{NRQED(\texorpdfstring{$\mu p$}{})}
\label{NRQED(mu)}

In the muon-proton sector, by integrating out the $m_\pi \sim m_{\mu}$ scale, an
EFT for non-relativistic muons and protons, relativistic electrons and photons appears. In
principle, we should also consider neutrons but they play no role at
the precision we aim. The effective theory has a hard
cut-off $\nu \ll m_\pi$ and therefore pion and Delta particles have been
integrated out. The effective Lagrangian reads
\be
{\mathcal L}_{\rm NRQED(\mu)}=
{\mathcal L}_{\gamma}
+
{\mathcal L}_{e}
+
{\mathcal L}^{(\rm NR)}_{\mu}
+
{\mathcal L}_{N}
+
{\mathcal L}_{Ne}
+
{\mathcal L}_{N\mu}^{(\rm NR)}
\,.
\ee
The pure photon sector is approximated by the following Lagrangian
\be
\label{Lg}
{\mathcal L}_\gamma=-\frac{1}{4}F^{\mu\nu}F_{\mu \nu} 
+
\left(\frac{d_{2}^{(\mu)} }{ m_{\mu}^2}+\frac{d_{2}}{ m_p^2}+\frac{d_{2}^{(\tau)} }{ m_{\tau}^2} 
\right)
F_{\mu \nu} D^2 F^{\mu \nu}
\,,
\ee
$d_{2}^{(\mu)}$ and $d_{2}^{(\tau)}$ are generated by the vacuum polarization loops with only muons and taus respectively. 
At ${\mathcal O}(\al)$ they read
\be
 d_{2}^{(\mu)}=\frac{Z_{\mu}^2\alpha}{ {60\pi}}+{\mathcal O}(\al^2) \,,
 \qquad 
 d_{2}^{(\tau)}=\frac{\alpha}{ {60\pi}}+{\mathcal O}(\al^2)
 \label{d2}
\,.
\ee
 
The hadronic effects of the vacuum polarization are encoded in $d_{2}$:
\be
d_{2}=\frac{m_p^2 }{4}\Pi^\prime_{h}(0)=\frac{Z_p^2\al}{60\pi}+d^{\rm had}_2+{\mathcal O}(\al^2)
\,.\label{d2p}
\ee
$\Pi^\prime_{h}(0)$ is the derivative of the hadronic vacuum polarization
(we have defined $\Pi_{h}(-{\bf k}^2)=-{\bf k}^2\Pi^\prime_{h}(0)+\,.\,.\,.$). The 
experimental figure for the total hadronic contribution reads
$\Pi_{h}^{\prime} \simeq 9.3 \times \, 10^{-3} \, {\rm GeV}^{-2}$ 
\cite{Jegerlehner:1996ab}. 
Following standard practice, we have singled out the contribution due to the loops of protons (assuming them to be point-like) in the second equality of Eq.~(\ref{d2p}). Note though that $d_2^{\rm had}$ is still of order $\al$.

The electron sector reads ($i D_\mu=i\partial_\mu-eA_\mu$)
\be
\label{Ll}
{\mathcal L}_e= \bar l_e  (i\dsl-m_{l_e}) l_e
\,.
\ee
We do not include the term 
\be
-\frac{e g_{l_e}
}{ m_{\mu}}\bar l_e \sigma_{\mu\nu}l_eF^{\mu\nu}
\,,
\ee
since the coefficient $g_{l_e}$ is suppressed by powers of $\al$ and 
the mass of the lepton. Therefore, it would give contributions beyond the
accuracy we aim. In any case, any eventual contribution would be
absorbed in a low energy constant.

The muonic sector reads
\bea
{\mathcal L}^{(\rm NR)}_{\mu}&=& 
l^\dagger_{\mu} \Biggl\{ i D_{\mu}^0 + \, \frac{{\bf D}_{\mu}^2}{ 2 m_{\mu}} +
 \frac{{\bf D}_{\mu}^4}{
8 m^3_{\mu}} + e \frac{c_F^{(\mu)} }{ 2m_\mu}\, {\bf \bfsigma \cdot B}\nn
 \\
 &&
 + 
e\frac {c_D^{(\mu)} }{ 8m_{\mu}^2}
   \left[{\bf \bfnabla \cdot E }\right]
   + 
 i e \frac{c_S^{(\mu)} }{ 8 m^2_{\mu}}\,  {\bf \bfsigma \cdot \left(D_{\mu} \times
     E -E \times D_{\mu}\right) } 
\Biggr\} l_{\mu},
\eea
with the following definitions: $ i D^0_\mu=i\partial_0 -Z_\mu eA^0$,
$i{\bf D}_\mu=i{\bfnabla}+Z_\mu e{\bf A}$ and $Z_\mu=1$. The Wilson
coefficients can be computed order by order in $\al$. They read (where we have used the fact that  
$c_S^{(\mu)}=2c_F^{(\mu)}-Z_{\mu}$ \cite{Manohar:1997qy})
\begin{eqnarray}
c_F^{(\mu)}&=&Z_{\mu}\left(1+\frac{Z_{\mu}^2\al}{2\pi} +{\mathcal O}(\al^2)\right) \,,
\label{cfi}\\
c_S^{(\mu)}&=&Z_{\mu}\left(1+\frac{Z_{\mu}^2\al}{\pi} +{\mathcal O}(\al^2)\right) \label{csi}\,.
\end{eqnarray}

Taking the values of the form factors for the muon-electron difference computed in \cite{Barbieri:1973kk} and those for the electron computed in \cite{Barbieri:1972as}, we can deduce the following expression for the 
$c_{D,\MS}^{(\mu)}(\nu)$ Wilson coefficient\footnote{In NRQED($\mu p$), the electron has not been integrated out. Therefore, 
Eq.~(\ref{cDmu}) is not the $c_D^{(\mu)}$ Wilson coefficient of NRQED($\mu p$). 
Eq.~(\ref{cDmu}) will show up after lowering the muon energy cut-off below the electron mass in pNRQED. Still we choose to present it here as otherwise we would be forced to do an extra intermediate 
matching computation that it is unnecessary to obtain the final result. Since we have integrated out the 
electron, note also that $\al=1/137.14...$ in this equation, i.e. any running associated to the electron is written explicitly in Eq.~(\ref{cDmu}).}:

\bea
\label{cDmu}
c_{D,\MS}^{(\mu)}(\nu)&=&
Z_{\mu}
\left(
1+\frac{4\alpha}{3\pi}Z_{\mu}^2\ln\left(\frac{m_\mu^2}{\nu^2}\right)
\right.
\\
&&+\left(\frac{\alpha}{\pi}\right)^2Z_{\mu}^2
\left\{
\frac{8}{9} \ln^2\left(\frac{m_\mu}{m_e}\right)
-\frac{40}{27} \ln\left(\frac{m_\mu}{m_e}\right)+\frac{85}{81}+ \frac{4\pi ^2}{27}
\right.
\nn\\
&&
\left.\left.
+Z_{\mu}^2
\left[\frac{\pi ^2 }{6}\left(18 \ln(2)-\frac{40}{9}\right)
-\frac{1523}{324}
-\frac{9 }{2}\zeta(3)
\right]
+{\mathcal O}\left(\frac{m_e}{m_{\mu}}\right)\right\}
\right)
\nn\\
\nn
&+&{\mathcal O}\left(\al^3\right)
\,.
\eea

Note that written in this way one can easily read the  ${\mathcal O}(\als^2)$ $C_f^2$ and $C_fT_Fn_l$ (for the case of massive $n_l$ quarks) coefficients that would appear in the analogous Wilson coefficient $c_D$ in QCD. The second line in 
Eq.~(\ref{cDmu}) would correspond to the $C_fT_Fn_l$  term and the 
third line to the $C_f^2$ one. 

For the Lamb shift computation we perform in this paper we only need $c_D^{(\mu)}$ with ${\mathcal O}(\al^2\times \ln)$ accuracy. We also include the finite piece for completeness but neglect ${\mathcal O}(m_e/m_{\mu})$ terms.
 Note that analogous ${\mathcal O}(\al^2)$ terms (changing $m_{\mu}$ by $m_p$ and either keeping 
$m_e$ or changing it by $m_{\mu}$) would exist for $c_D^{(p)}$ if computing the Wilson coefficient as if the proton were point-like at the $m_p$ scale. Even 
if these effects are small, they should be taken into account for eventual comparisons with lattice where typically only the hadronic correction is computed.

For the proton sector we have 
\bea
\label{LNdelta}
 {\mathcal L}_{N}&=& N^\dagger_{p} \Biggl\{iD_0+ \frac{{\bf
D}^2_p}{2 m_{p}} + 
\frac {{\bf D}_p^4}{
8 m^3_{p}} - e \frac{c_F^{(p)} }{ 2m_p}\, {\bf \bfsigma \cdot B}\nn
\\
&& 
-e\frac{c_D^{(p)} }{ 8m_p^2}
   \left[{\bf \bfnabla \cdot E }\right] 
 - ie \frac{ c_S^{(p)} }{8m_p^2}\, \bfsigma \cdot \left({\bf D}_p
     \times {\bf E} -{\bf E}    \times {\bf D}_p\right) 
\Biggr\} N_{p}
\,,
\eea
where $ i D^0_p=i\partial_0 +Z_peA^0$, $i{\bf D}_p=i{\bfnabla}-Z_pe{\bf A}$ and for the proton $Z_p=1$. The proton Wilson coefficients are hadronic, non perturbative quantities. 
In some cases they can be directly related with low energy constants, for instance with the anomalous magnetic moment of the proton, $\kappa_p=1.792847356(23)$ 
\cite{Agashe:2014kda}:
\begin{eqnarray}
c_F^{(p)}&=&Z_p+\kappa_p =Z_p+\kappa_p^{\rm had}+\frac{Z_p^3\al}{2\pi}+{\mathcal O}(\al^2)
\label{cfp},\\
c_S^{(p)}&=&Z_p+2\kappa_p  =Z_p+2\kappa_p^{\rm had}+\frac{Z_p^3\al}{\pi}+{\mathcal O}(\al^2)\label{csp}
\,.
\end{eqnarray}
Note that $\kappa_p$ includes ${\mathcal O} (\alpha)$ effects. In principle, this is also so for $\kappa_p^{\rm had}$, to which we have 
subtracted the proton-associated point-like contribution to the anomalous magnetic moment 
(note that the point-like result is a bad approximation for $c_F^{(p)}$, even though it gives the right order of magnitude). 
The case of $c_D^{(p)}$ is more complicated (a more detailed discussion can be found in Ref.~\cite{Pineda:2004mx}). 
It can be written in the following way in terms of the electromagnetic current form factors at zero momentum 
($F_1(0)=Z_p$):
\be
c^{(p)}_D=Z_p+2F_2(0)+8F_1^{\prime}(0)=Z_p+8m_p^2\left.\frac{d G_{p,E}(q^2) }{ d\,q^2}\right|_{q^2=0}
\,.
\ee
 This object is infrared divergent, which makes it scale and 
scheme dependent. 
This is not a problem from the EFT point of view but makes the 
definition of the proton radius ambiguous. 
The standard practice is to make explicit the proton-associated point-like contributions to the computation. 
In practice this means that one uses the following definition for the proton radius
\be
c_{D,\MS}^{(p)}(\nu)\equiv Z_p+\frac{4}{3}\frac{Z_p^3\alpha}{\pi}\ln\left(\frac{m_p^2}{\nu^2}\right)
+\frac{4}{3} r_p^2 m_p^2+{\mathcal O}(\al^2).
\ee
In other words (up to ${\mathcal O}(\al^2)$ corrections)
\be
c_{D,\MS}^{(p)}(m_p)-Z_p\equiv \frac{4}{3} r_p^2 m_p^2\,.
\ee
Note that $r_p$ includes ${\mathcal O}(\al)$ terms in its definition. This should be kept in mind when comparing with lattice determinations. 
 Note also that it is not natural to set $\nu=m_p$, or, in other words, to assume that the proton is point-like up to (and beyond) the scales of the proton mass;
$\frac{4}{3} r_p^2 m_p^2 \simeq 21.3$, to be compared with "1" for a point-like particle. This illustrates that the point-like result does not even give the right order of
magnitude of $c_D$\footnote{Although not directly relevant for the specific computation of this paper, note that this 
also happens for the Wilson coefficients $c_{A_1}$ and $c_{A_2}$ (for the definition see Ref.~\cite{Pineda:2004mx}), for which their physical values are far from zero: $c_{A_1}\simeq 12$ and $c_{A_2}\simeq -72$, even though for a point-like particle their values would be "1" and "0" respectively (up to $\mathcal O(\al)$ corrections).}.
 
 ${\mathcal L}_{Ne}$ refers to the four-fermion operator made of nucleons and
 (relativistic) electrons. It will not contribute to the spectrum at 
 ${\mathcal O}(m_r\al^5)$.
 Therefore, we will not consider it any further. For a more detailed discussion see Ref.~\cite{Pineda:2002as}. 
 
Finally, we consider the four-fermion operators\footnote{The coefficients $c_{3}$ 
and $c_{4}$ should actually read $c_{3}^{pl_\mu}$ and $c_{4}^{pl_\mu}$, as they actually depend on the
nucleon and lepton the four-fermion operator is made of. Nevertheless, to ease the notation we eliminate those 
indices.}:
\be
{\mathcal L}_{N\mu}^{\rm NR}=
\displaystyle\frac{c_{3}}{m_p^2} N_p^{\dagger}
  N_p \ {l}^{\dagger}_\mu l_\mu
-\displaystyle\frac{c_{4}}{m_p^2} N_p^{\dagger}{\bfsigma}
  N_p \ {l}^{\dagger}_\mu{\bfsigma} l_\mu
\,.
\ee
Again in this case it is common practice to single-out the proton-associated point-like contribution. Note that this assumes that one can treat the proton as point-like at energies of the order of the proton mass. We have already seen that this is a bad approximation for $c_D$ and other Wilson coefficients. Nevertheless, we keep this procedure for the sake of 
comparison. Therefore, 
\bea
c_{3}(\nu)&\equiv&-\frac{m_p}{m_{\mu}}d_s(\nu)+c_{3}^{\rm had}+{\mathcal O}(\al^3)
\,,
\\
c_{4}&\equiv&-\frac{m_p}{m_{\mu}}d_v+c_{4}^{\rm had}+{\mathcal O}(\al^3)
\,,
\eea
where the point-like Wilson coefficients read as follows:
\begin{eqnarray}
d_s(\nu)&=&-\frac{Z^2\alpha^2}{m_\mu^2-m_p^2}\left[m_\mu^2\left(\ln \frac{m_p^2}{\nu^2}+\frac{1}{3}\right)-m_p^2\left(\ln \frac{m_\mu^2}{\nu^2}+\frac{1}{3}\right)\right],\\
d_v&=&\frac{Z^2\alpha^2}{m_\mu^2-m_p^2}m_\mu m_p \ln\frac{m_\mu^2}{m_p^2}.
\end{eqnarray}
The expression of $d_s$ should be understood in the $\MS$ scheme, 
$d_v$ on the other hand is finite. 
$d_s$ was computed in Ref.~\cite{Pineda:1998kj} and $d_v$  in Ref.~\cite{Caswell:1985ui}.

$c^{\rm had}_3$ encodes all the hadronic effects to the spin-independent four-fermion Wilson coefficient. 
At ${\mathcal O}(\al^2)$ it is generated by the two-photon exchange contribution. Since $c_3^{\rm had}$ depends linearly on the muon mass, it is dominated by the infrared dynamics and diverges linearly in the chiral limit. This produces an extra
$m_{\mu}/m_{\pi}$ suppression with respect to its natural size, and allows us to compute the leading 
pure-chiral and Delta-related effects in a model independent way. The complete matching computation 
between HBET and NRQED was made in Ref.~\cite{Peset:2014jxa} to which we refer for details (partial results can be found in \cite{Pineda:2004mx,Nevado:2007dd}, and in Ref.~\cite{Alarcon:2013cba} in the context of relativistic baryon effective theory). Overall we obtained 
\bea
\nn
c^{\rm had}_{3} &\sim& \al^2 \frac{m_{\mu}}{m_{\pi}}\left[1+\#\frac{m_{\pi}}{\Delta}+\cdots\right]
+{\mathcal O}\left(\al^2 \frac{m_{\mu}}{\lQ}\right)
\\
&=&\al^2 \frac{m_{\mu}}{m_{\pi}} \begin{cases}\displaystyle 
47.2(23.6)&(\pi),\\
56.7(20.6)&(\pi+\Delta), \end{cases}
\label{eq:c3had}
\eea
where the upper and lower numbers refer to the matching computation with only pions, or with pions and the Delta particle, respectively. For comparison, the value $c^{\rm had}_{3}=\al^2 \frac{m_{\mu}}{m_{\pi}}54.4(3.3)$, 
which follows from the analysis in Ref.~\cite{Birse:2012eb}, was used in Ref.~\cite{Antognini:1900ns}. 
We refer to Ref.~\cite{Peset:2014jxa} for a detailed discussion on the status of these hadronic determinations and focus on the QED-like computations in this paper. 

$c^{\rm had}_4$ encodes all the hadronic effects to the spin-dependent four-fermion Wilson coefficients. 
As in the previous case, this coefficient diverges in the chiral limit. Nevertheless, it 
 only does so logarithmically (unlike 
in the previous case, where the divergence was linear). Such computation can be found in Ref.~\cite{Pineda:2002as}. Still it 
is possible to determine $c^{\rm had}_4$ from the analogous one of the proton-electron four-fermion 
operator determined in Ref.~\cite{Pineda:2002as}. This was done in Ref.~\cite{Peset:2014jxa}, from where we quote the result
\be
c^{\rm had}_{4} \simeq-46\al^2 
\,.
\ee
 
\section{pNRQED}
\label{Sec:pNRQCD}
After integrating out scales of ${\mathcal O}(m_{\mu}\al \sim m_e)$, 
the resulting effective theory is pNRQED~\cite{Pineda:1997bj}. 
This EFT naturally gives a Schr\"odinger-like formulation of the bound-state problem but still keeping the quantum field theory nature of the interaction with ultrasoft photons, as well as keeping the information due to high energy modes (of a quantum field theory nature) in the Wilson coefficients of the theory. pNRQED has been applied to hydrogen \cite{Pineda:1997ie}, positronium~\cite{Pineda:1998kn} and muonic 
hydrogen~\cite{Pineda:2002as,Pineda:2004mx} providing with much of the information needed for this paper. In particular in the last reference the explicit form of the Lagrangian was presented (up to ${\mathcal O}(m_r\al^5)$). We repeat it here but generalized to the case of arbitrary charges:
\bea
&&L_{\rm pNRQED} =
\int d^3{\bf x} d^3{\bf X} S^{\dagger}({\bf x}, {\bf X}, t)
                \Biggl\{
i\partial_0 - \frac{ {\bf p}^2 }{2m_{r}} + \frac{ {\bf p}^4 }{ 8m_{\mu}^3}+ 
\frac{ {\bf p}^4 }{ 8m_p^3} - \frac{ {\bf P}^2}{ 2M}
\\
&&
\nonumber
- V ({\bf x}, {\bf p}, {\bfsigma}_1,{\bfsigma}_2) + e 
\left(
\frac{Z_{\mu} m_p+Z_p m_{\mu}}{m_p+m_{\mu}}
\right)
{\bf x} \cdot {\bf E} ({\bf X},t)
\Biggr\}
S ({\bf x}, {\bf X}, t)- \int d^3{\bf X} \frac{1}{ 4} F_{\mu \nu} F^{\mu \nu}
\,,
\eea
where $M=m_{\mu}+m_p$, $m_r= \frac{m_{\mu}m_p }{ m_{\mu}+m_p}$, ${\bf x}$ and 
${\bf X}$, and
${\bf p}$ and ${\bf P}$ are the relative and center of mass coordinate and momentum
respectively.

$V$ can be written as an expansion in $1/m_{\mu}$, $1/m_p$, $\al$, ... 
We will assume $1/r \sim m_e$ (which is realistic for the case at hand) 
and that $m_{\mu} \ll m_p$. We then organize the potential as an expansion in $1/m_{\mu}$:
\be
V ({\bf x}, {\bf p}, {\bfsigma}_1,{\bfsigma}_2)
=
V^{(0)}(r)+{V^{(1)}(r) }+{V^{(2)}(r) }+\cdots\,,
\ee
where
\be
V^{(n)} \propto \frac{1}{m_{\mu}^n}.
\ee
We will also make the expansion in powers of $\al$ explicit. This means that 
\be
V^{(n,r)} \propto \frac{1}{m_{\mu}^n}\al^r.
\ee

$V^{(0,1)}=-\frac{Z\al}{r}$ has to be included exactly in the leading order Hamiltonian to yield the leading-order solution 
to the bound-state problem:
\be
h=\frac{{\bf p}^2}{2m_r}-\frac{Z\al}{r}
\,.
\ee
Thus, the contribution to the energy of a given potential is 
$$
\langle {V^{(n,r)} }\rangle \sim m_{\mu}\al^{1+n+r}
$$ 
up to large logarithms or potential suppression factors due to powers of $1/m_p$. Iterations of the potential are dealt with using standard quantum mechanics perturbation theory producing corrections such as:
\be
\langle V^{(n,r)}\cdots V^{(m,s)} \rangle \sim m_{\mu}\al^{1+n+r+(1+m+s-2)}
\ee
and alike. Therefore, in order to reach the desired ${\mathcal O}(m\al^5)$ accuracy, $V^{(0)}$ 
has to be computed up to $\mathcal O(\al^4)$, $V^{(1)}$ up to $\mathcal O(\al^3)$, 
$V^{(2)}$ up to $\mathcal O(\al^2)$ and $V^{(3)}$ up to $\mathcal O(\al)$. 

\subsection{The static potential: \texorpdfstring{$V^{(0)}$}{}}

The Fourier transform of $V^{(0)}$ reads
\begin{equation}
{\tilde V}^{(0)} \equiv  - 4\pi Z \frac{\alpha_{\tilde V}(k)}{ {\bf k}^2}
\equiv \sum_{n=1}^{\infty}\tilde V^{(0,n)},  
\label{defpot0}
\end{equation}
which in fact defines $\alpha_{\tilde V}$, the coupling constant associated to the static potential, 
which is gauge invariant. The contribution associated to the electron vacuum polarization ($\Pi(0)=0$)
$$
\Pi(k^2)=\al\Pi_{1}(k^2)+\al^2\Pi_{2}(k^2)+\al^3\Pi_{3}(k^2)+...
$$ 
provides with another very popular definition for the effective coupling that enjoys the nice properties of 
gauge invariance and scheme/scale independence:
\be
\alpha_{\rm eff}(k)=\al\frac{1 }{ 1+\Pi(-{\bf k}^2)}=\al-\frac{\al^2}{\pi}\Pi_{1}+\frac{\al^3}{\pi^2}(\Pi_{1}^2-\Pi_{2})
+\frac{\al^4}{\pi^3}(-\Pi_1^3+2\Pi_1\Pi_2-\Pi_{3})+{\mathcal O}(\al^5)
\,.
\ee
$\alpha_{\rm eff}$ corresponds to Dyson summation. If we express
$\alpha_{\tilde V}(k)$ in terms of $\alpha_{\rm eff}(k)$, we have 
\be 
\alpha_{\tilde V}(k)=
\alpha_{\rm eff}(k)+\sum_{\genfrac{}{}{0pt}{}{n,m=0}{ n+m=\text{even}>0}} Z_{\mu}^nZ_p^m\al_{\rm eff}^{(n,m)}(k) 
\equiv \alpha_{\rm eff}(k)+\delta \al(k)\,, \quad \quad \delta \al(k)=\mathcal O(\al^4)
.
\ee
\begin{equation}
{\tilde V}^{(0,1)} \equiv  - 4\pi Z \frac{\alpha  }{ {\bf k}^2},  
\label{defpot01}
\end{equation}
is nothing but the Coulomb potential.  In order to achieve ${\mathcal O}(m_r\al^5)$ accuracy we need to know
$\Pi^{(1)}$, $\Pi^{(2)}$, $\Pi^{(3)}$ and the leading, non-vanishing,
contributions to $\al_{\rm eff}^{(2,0)}(k)$, $\al_{\rm eff}^{(0,2)}(k)$ and
$\al_{\rm eff}^{(1,1)}(k)$. 

\begin{figure}[h]
  \begin{center}
   \includegraphics[width=0.25\textwidth]{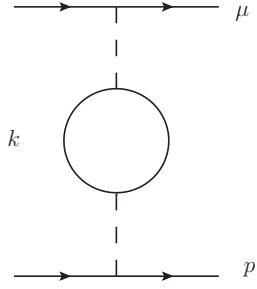}
  \end{center}
  \caption{\it One-loop electron vacuum polarization contribution to the static potential.}
  \label{fig:VP}
\end{figure}

The next-to-leading order term of the static potential is displayed in fig. 1 and reads
\be
\tilde V^{(0,2)}_{\rm VP}(k)= 4\pi Z\frac{\alpha^2}{\pi}\frac{\Pi_1(-{\bf k}^2) }{ {\bf k}^2},
\ee
where
\begin{equation}
\Pi_{1}(k^2)=k^2\int_4^\infty d q^2\frac{1}{{q^2(m_e^2q^2-k^2)}}u(q^2),
\end{equation}
and
\be
u(q^2)=\frac{1 }{ 3}\sqrt{1-\frac{4 }{ q^2}}\left(1+\frac{2 }{ q^2}\right)
\,.
\ee
Thus, we may write for the potential in position space
\begin{equation}
V^{(0,2)}_{\rm VP}(r)=-\frac{Z\alpha}{ r}\frac{\alpha}{\pi}\int_4^\infty  \frac{dq^2}{q^2} u(q^2) e^{-2m_e r q}.
\end{equation}

\begin{figure}[h]
  \begin{center}
   \includegraphics[width=\textwidth]{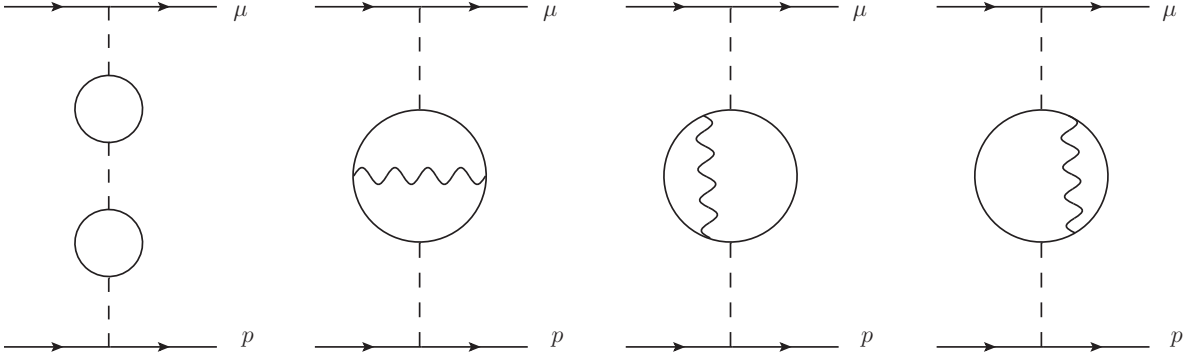}
  \end{center}
  \caption{\it Diagrams contributing to $V^{(0,3)}$.}
  \label{fig:2loopVP}
\end{figure}

The next-to-next-to-leading order term of the static potential is produced by the diagrams depicted in Fig.~\ref{fig:2loopVP}, which can be 
understood as a correction to the vacuum polarization. It was computed by K\"allen and Sabry \cite{Kallen:1955fb} 
and reads
\be
\tilde V^{(0,3)}_{\rm VP}(k)= 4\pi Z \frac{\alpha^3}{\pi^2}\frac{\Pi^2_1(-{\bf k}^2)-\Pi_2(-{\bf k}^2) }{ {\bf k}^2},
\ee
\begin{equation}
\Pi^2_1(k^2)-\Pi_2(k^2)=k^2\int_4^\infty d q^2\frac{1}{{q^2(m_e^2q^2-k^2)}}u^{(2)}(q^2),
\end{equation}
where
\bea
u^{(2)}(q^2)&=&\frac{1}{3} \left[\tau\left(-\frac{19}{24}+\frac{55 }{72}\tau^2-\frac{1}{3}\tau^4-\frac{3-\tau^2}{2}\ln\left(\frac{64 \tau^4}{\left(1-\tau^2\right)^3}\right)\right)
\right.
\nn
\\
&&
+\ln\left(\frac{1+\tau}{1-\tau}\right) \left(\frac{33}{16}+\frac{23}{8} \tau^2-\frac{23}{16} \tau^4
+\frac{1}{6} \tau^6+\left(\frac{3}{2}+\tau^2-\frac{\tau^4}{2}\right) \ln\left(\frac{(1+\tau)^3}{8\tau^2}\right)\right)
\nn
\\
&&
\left.
+
\left(3+2\tau^2-\tau^4\right) 
\left(2 \text{Li}_2\left(\frac{1-\tau}{1+\tau}\right)+\text{Li}_2\left(\frac{-1+\tau}{1+\tau}\right)\right)
\right],
\eea
with
\bea
{\rm Li}_2(x)=-\int_0^z du\frac{\ln(1-u)}{u},\;\;z\in \mathbb{C}\,\backslash [1,\infty) \qquad {\rm and} \qquad \tau=\sqrt{1-\frac{4}{q^2}}.
\eea

The next-to-next-to-next-to-leading order term of the static potential coming from the vacuum polarization reads
\be
\tilde V^{(0,4)}_{\rm VP}(k)= 4\pi Z\frac{\alpha^4}{\pi^3}\frac{-\Pi_1^3(-{\bf k}^2)+2\Pi_1(-{\bf k}^2)\Pi_2(-{\bf k}^2)-\Pi_{3}(-{\bf k}^2) }{ {\bf k}^2}.
\ee
This object (more specifically $\Pi_{3}$) has been computed in Ref.~\cite{Kinoshita:1979dy}, see also \cite{Kinoshita:1998jf} where the complete set of 
diagrams can be found.

\begin{figure}[h]
  \begin{center}
  \includegraphics[width=0.75\textwidth]{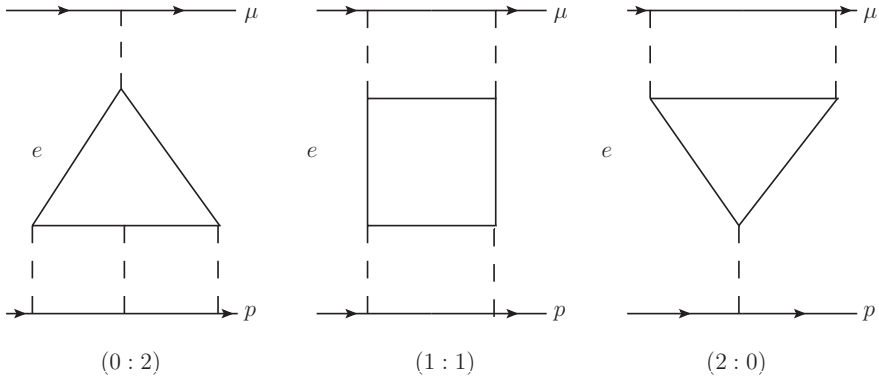}
  \end{center}
  \caption{\it Light-by-light contribution to the static potential. The first and third diagram are the contributions to 
  $\al_{\rm eff}^{(2,0)}(k)$ and $\al_{\rm eff}^{(0,2)}(k)$ respectively. The second diagram contributes to 
  $\al_{\rm eff}^{(1,1)}(k)$.}
  \label{fig:lbl}
\end{figure}

The remaining next-to-next-to-leading order contribution to the static potential is generated by diagrams that cannot be completely associated to the vacuum polarization, and is encoded in $\delta \al(k)$.
Its sum is constrained to fulfil $n+m=\text{even}$ because of the Furry theorem.
Each $\al_{\rm eff}^{(n,m)}(k)$ is also gauge invariant. The leading, non-vanishing, contributions are 
$\al_{\rm eff}^{(2,0)}(k)$, $\al_{\rm eff}^{(0,2)}(k)$ and
$\al_{\rm eff}^{(1,1)}(k)$. They have an expansion in $\al$ themselves. Since each of them is ${\mathcal O}(\al^4)$, we can approximate them by its leading order expression, which is produced by the light-by-light diagrams displayed in 
Fig.~\ref{fig:lbl}. This object could be deduced from the computation in Ref.~\cite{Karshenboim:2010cq}
(we truncate the $\al_{\rm eff}$ expressions to its leading order)
\be 
\label{VLbL}
\tilde V^{(0,4)}_{\rm LbL}(k)= 
-\frac{4\pi Z}{\bf k^2} \left((Z^2_{\mu}+Z^2_p)\al_{\rm eff}^{(2,0)}(k) +Z_{\mu}Z_p\al_{\rm eff}^{(1,1)}(k)\right)
,
\ee
where we have already used that $\al_{\rm eff}^{(2,0)}(k)=\al_{\rm eff}^{(0,2)}(k)$.

\subsection{The potential beyond the static limit}

In the matching scheme used in this paper (off-shell in the Coulomb gauge) the $1/m$ potential is  zero in QED without light fermions (see Ref.~\cite{Pineda:1998kn}). This is no longer so after the
 inclusion of light fermions (the electron) 
into the computation. Yet, after inspection of the diagrams that may contribute, they would produce, at most,   
${\mathcal O}(m_r\al^6)$ corrections to the energy, so they will
be neglected in the following.

\begin{figure}[h]
  \begin{center}
  \includegraphics[width=0.55\textwidth]{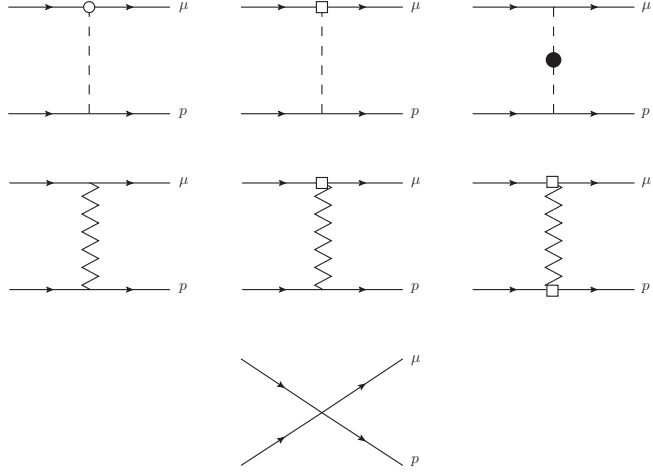}
  \end{center}
\caption{\it The non-zero relevant diagrams for the matching at tree level
       in the Coulomb gauge. The dashed and zigzag lines represent the
       $A_0$ and ${\bf A}$ fields respectively, while the continuous
       lines represent the fermion and antifermion fields. For the $A_0$ the circle is the vertex
       proportional to $c_D$, the square to $c_S$ (spin dependent) and
       the black dot to $d_2$, while for
      ${\bf A}$ the square is the vertex proportional to $c_F$ and
       the other vertex appears from the covariant derivative in the
       kinetic term. The last diagram is proportional to $c_3$ and
       $c_4$. The symmetric diagrams are not displayed. It corresponds to Eq.~(\ref{VbcD})}
\label{figtree}
\end{figure}

\begin{figure}[h]
  \begin{center}
  \includegraphics[width=0.55\textwidth]{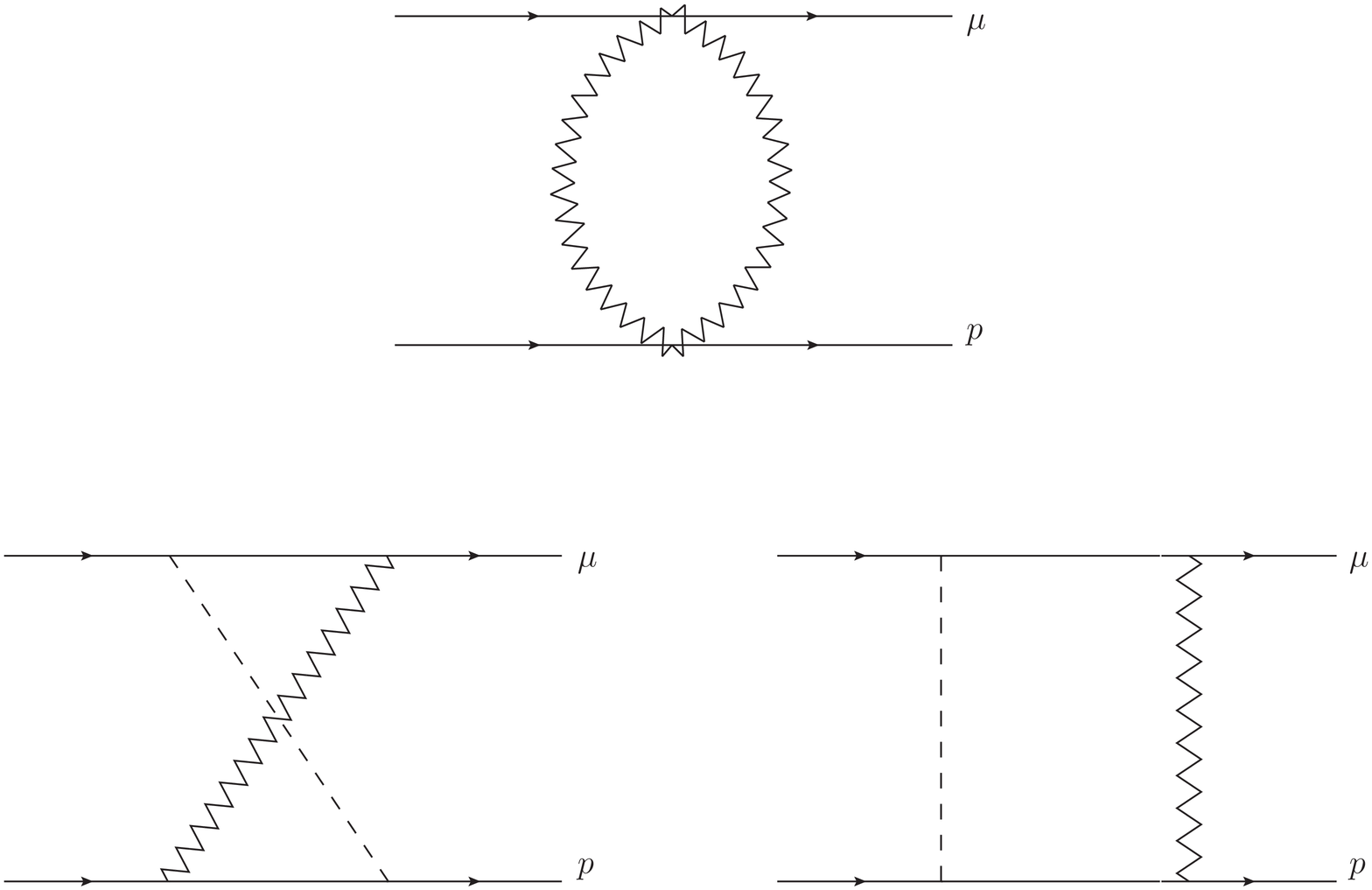}
  \end{center}
\caption{\it The non-zero relevant diagrams for the matching at one loop
       in the Coulomb gauge. The dashed and zigzag lines represent the
      $A_0$ and ${\bf A}$ fields respectively, while the continuous
       lines represent the fermion and antifermion. The interactions
       for ${\bf A}$ are the ones which appear from the covariant space
       derivatives in the kinetic term,
       while for $A_0$ comes from the covariant time derivative. The
       symmetric diagrams are not displayed. They correspond to Eq.~(\ref{tildeV1loop}).}
\label{fig1loop}
\end{figure}

\begin{figure}[h]
  \begin{center}
  \includegraphics[width=0.25\textwidth]{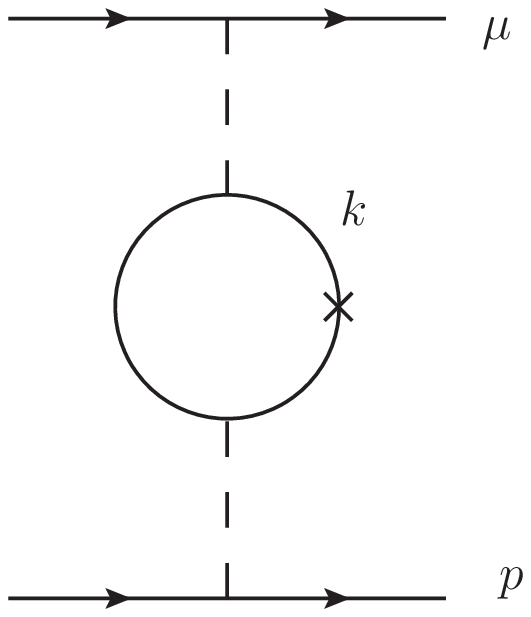}
  \end{center}
\caption{\it Symbolic representation of the leading correction to the static 
potential due to the Taylor expansion of the electron vacuum
  polarization in the Coulomb gauge in powers of $k^0=E_1-E_1^{\prime}$. It corresponds to Eq.~(\ref{dtv}).}
\label{figvacuum}
\end{figure}

At  order $1/m^2$ the expression of the potential in momentum space was obtained in 
Ref.~\cite{Pineda:2004mx}. We summarize its different contributions here. 
If we switch off the electron vacuum polarization effects, the computation 
would correspond to the muonium case (or positronium for the equal mass case). The relevant diagrams 
in such situation are presented in Figs.~\ref{figtree} and \ref{fig1loop} (following the classification of 
Ref. \cite{Pineda:1998kn} generalized to the non-equal mass case). The one-loop diagrams in Fig.~\ref{fig1loop}
produce the potential quoted in Eq.~(\ref{tildeV1loop}) (in the $\MS$ scheme). The tree-level 
diagrams of Figs.~\ref{figtree} produce the potential quoted in Eq.~(\ref{VbcD}) with $\al_{\rm eff} = \al$. 
In order to incorporate the electron vacuum polarization effects at one loop we replace 
$\al \rightarrow \al_{\rm eff} \simeq \al -\al^2\Pi_1(-{\bf k}^2)$ in Eq.~(\ref{VbcD}). This means including the vacuum polarization in the 1st, 2nd, 4th, 5th and 6th diagram in Fig.~\ref{figtree}. On top of that one has to include 
the contribution coming from Fig.~\ref{figvacuum}, which appears from the 
Taylor expansion in powers of the transfer energy of the vacuum polarization when doing the matching computation off-shell (for further details see the discussion in Ref.~\cite{Pineda:2004mx}). It produces the potential quoted in 
Eq.~(\ref{dtv}). Overall, the complete expression 
for the $1/m_{\mu}^2$ potential at one loop is summarized in Eqs.~(\ref{VbcD}), (\ref{tildeV1loop}) and (\ref{dtv}), which we list now

\bea
{\tilde V}^{(2)}_{\rm tree+VP} &=&\frac { \pi\al_{\rm eff}(k) }{ 2} \left(
  Z_p\frac{c_D^{(\mu)}}{ m_{\mu}^2}+Z_{\mu}\frac{c_D^{(p)} }{ m_p^2}
\right)
\nn\\
 && -  i 2 \pi\al_{\rm eff}(k)  \frac{ ({\bf p} \times
  {\bf k}) }{{\bf k}^2}\cdot \left( Z_p\frac{c_S^{(\mu)} {\bf S}_{\mu}}{ m_{\mu}^2} + 
Z_{\mu}\frac{c_S^{(p)} {\bf S}_p }{ m_p^2}
\right)
\nn\\
&&- Z 16\pi\al\left(\frac{ d_2^{(\mu)} }{ m_{\mu}^2}
+\frac{ d_2^{(\tau)} }{ m_{\tau}^2}+\frac{ d_{2} }{ m_p^2}
\right)
\nn\\
&& - Z\frac{ 4 \pi\al_{\rm eff}(k) }{m_{\mu}m_p} \left( \frac{ {\bf p}^2
  }{ {\bf k}^2} - \frac{({\bf p} \cdot {\bf k})^2 }{ {\bf k}^4}
\right)
\nn\\
&& -\frac { i 4 \pi\al_{\rm eff}(k)}{ m_{\mu}m_p} \frac{ ({\bf p} \times
  {\bf k}) }{{\bf k}^2}\cdot(Z_pc_F^{(\mu)} {\bf S}_{\mu}+Z_{\mu}c_F^{(p)} {\bf S}_p)
\nn\\
&& { 4 \pi\al_{\rm eff}(k) c_F^{(\mu)}c_F^{(p)} }{ m_{\mu}m_p} 
\left( {\bf S}_{\mu}  \cdot {\bf S}_p - \frac{ {\bf S}_{\mu} \cdot {\bf k} {\bf
      S}_p \cdot {\bf k} }{ {\bf k}^2} 
\right)
\nn\\
&&
 -\frac{1}{ m_{p}^2}
\left(c_{3}-4c_{4} {\bf S}_{\mu}\cdot {\bf S}_p \right),
\,\label{VbcD}
\\
\label{tildeV1loop}
{\tilde V}^{(2,2)}_{\rm 1-loop} &=& \frac{ Z^2 \al^2}{ m_{\mu}m_p} 
\left(\frac{7}{3}\ln \frac{{\bf k}^2
    }{ \nu^2} + \frac{1 }{ 3} \right)
\,,
\\
{\tilde V}^{(2,2)}_{\rm off-shell} &=& -\frac{Ze^2 }{ 4 m_{\mu}m_p}
\frac{({\bf p}^2-{\bf p}^{\prime\,2})^2 }{ {\bf k}^2} \frac{\al }{ \pi} 
m_e^2 \int_4^{\infty}d(q^2)\frac{1 }{
  (m_e^2q^2+{\bf k}^2)^2}u(q^2)
\,,
\label{dtv}
\eea
where $\bf{S}_i=\bfsigma_i/2$ is the spin of the particle $i$. 
We stress again that Eq.~(\ref{tildeV1loop}) has been obtained in the $\MS$ scheme. 
The sum of these three potentials includes all terms of ${\mathcal O}(V^{(2,1)})$ and ${\mathcal O}(V^{(2,2)})$:
\be
{\tilde V}^{(2)}={\tilde V}^{(2)}_{\rm tree+VP} +{\tilde V}^{(2,2)}_{\rm 1-loop}+ {\tilde V}^{(2,2)}_{\rm off-shell} +{\mathcal O}({\tilde V}^{(2,3)})
\,.
\ee

\subsection{The potential in position space}

The matrix elements of the potentials that appear in the energy shifts are more efficiently computed in position space. Therefore, we also write the potentials in position space. In this case it is convenient to split the potential in a slightly different way than in momentum space. In particular, the vacuum polarization contributions are dealt with in an isolated way.
We follow the notation of  Ref.~\cite{Pineda:2004mx}. The contributions coming from tree-level diagrams read 

\begin{eqnarray}
\nn
 V_{\rm tree}^{(2)}&=&
 \frac{Z\alpha }{ {2m_\mu m_p}}\left[-\left\{\frac{1}{ r},\bf{p}^2\right\}
 +\frac{1}{ r^3}{\bf L}^2+4\pi \delta^{(3)}({\bf r})\right]-16\pi Z
 \alpha \left(\frac{d_2^{(\mu)}}{m_\mu^2}+\frac{d_2^{(\tau)}}{m_\tau^2}+\frac{d_2}{m_p^2}\right)\delta^{(3)}({\bf r})
 \\
 &&
 +
 \frac{\alpha }{ {2m_\mu m_p}}\left[ \left(\frac{{Z_{\mu}c_D^{(p)}m_\mu^2
 +Z_pc_D^{(\mu)}m_p^2}}{{m_\mu m_p}}
 \right)\pi \delta^{(3)}({\bf r})
 \right. \nonumber \\
&&+\left.Z_pc_F^{(\mu)}\frac{2 }{ r^3}{\bf L\cdot S}_\mu+Z_{\mu}c_F^{(p)}\frac{2 }{ r^3}{\bf L\cdot S}_p 
+m_\mu m_p\left\{\frac{Z_p c_S^{(\mu)}}{ m_\mu^2}\frac{1}{ r^3}{\bf L\cdot S}_\mu+\frac{Z_{\mu}c_S^{(p)}}{ m_p^2}\frac{1}{ r^3}{\bf L\cdot S}_p\right\}\right]\nonumber\\
 &&
 +
 \frac{\alpha }{ {2m_\mu m_p}}\left[ \frac{16\pi}{ 3} c_F^{(\mu)}c_F^{(p)} \delta^{(3)}({\bf r}) {\bf S}_\mu {\bf S}_p+\frac{c_F^{(\mu)}c_F^{(p)}}{ {2r^3}} \hat{S}_{p\mu}(\hat{\bf r})\right]
+\frac{1}{m_p^2}\left(-c_3+4{\bf S}_\mu{\bf S}_p c_4\right)\delta^{(3)}({\bf r}),
\nn
\\
\label{V2}
\end{eqnarray}
where $\hat{S}_{i j}(\hat{\bf r})=-4 ({\bf S_i \cdot S_j})+12(\bf {S_i\cdot \hat{r}})(\bf {S_j\cdot \hat{r}})$.

The Fourier transform of Eq. (\ref{tildeV1loop}) reads
\bea
\label{V221loop}
V^{(2,2)}_{\rm 1-loop}
=
 \frac{Z^2\al^2 }{ 3 m_p m_{\mu}}
 \Biggl[
 \delta^{(3)}({\bf x})
       ( 1- 7 \ln \nu^2 ) -
     \frac  { 7 }{ 2 \pi } {\rm reg} \frac{1 }{ {\vert {\bf x}
           \vert}^3}
           \Biggr].
\eea
Finally, the contributions associated to the one-loop vacuum polarization 
read\footnote{Note that the fourth line can be written in a way that makes the angular momentum structure more explicit:
\bea
&&
\int_4^\infty dq^2 \frac{u(q^2)}{q^2}
\left(\frac{\lambda^2 e^{-\lambda r}}{r}\left(1-\frac{\lambda r}{2}\right)
+ 2p^i \frac{ e^{-\lambda r}}{r}\left(\delta_{ij}+\frac{r_i r_j}{r^2}(1+\lambda r)\right)p^j\right)\\
\nn
&=&\int_4^\infty dq^2 \frac{u(q^2)}{q^2}
\left(2\left\{ {\bf p^2},\frac{e^{-\lambda r}}{r}\left(1+\frac{\lambda r}{2}\right)\right\}-2\frac{e^{-\lambda r}}{r^3}(1+\lambda r){\bf L^2}
+
\frac{\lambda^2}{r}e^{-\lambda r}\left(1+\frac{\lambda r}{2}\right)
-8\pi\delta^{(3)}(\vec r)\frac{}{}\right)
\,.
\eea
Nevertheless, one has to be careful when dealing with the right-hand-side of the equality, as the first and last term are separately divergent (but not their sum). 
}
\bea
&&
V^{(2)}_{\rm VP,1-loop}+V^{(2,2)}_{\rm off-shell}
=\frac{\alpha}{\pi}\int_4^\infty dq^2 \frac{u(q^2)}{q^2}
\\
&&
\nn
\times \left\{
\frac{\alpha}{8m_\mu^2 m_p^2}\left(Z_{\mu}c_D^{(p)}m_\mu^2+Z_pc_D^{(\mu)}m_p^2\right)
\left(4\pi \delta^{(3)}(\vec r)-\frac{\lambda^2 e^{-\lambda r}}{r}\right)
\right.
\\
\nn
&&
+
\frac{\alpha }{2}\left(Z_{\mu}c_S^{(p)}\frac{{\bf L\cdot S}_p}{m_p^2}+ Z_pc_S^{(\mu)}\frac{ {\bf L\cdot S}_\mu}{m_\mu^2} \right)
\left(\frac{ e^{-\lambda r}(1+\lambda r)}{r^3}\right)\\
&&
\nn
-\frac{Z_pZ_{\mu}\alpha}{4m_\mu m_p}
\left(\frac{\lambda^2 e^{-\lambda r}}{r}\left(1-\frac{\lambda r}{2}\right)
+ 2p^i \frac{ e^{-\lambda r}}{r}\left(\delta_{ij}+\frac{r_i r_j}{r^2}(1+\lambda r)\right)p^j\right)\\
&&
\nn
+\frac{\alpha }{m_\mu m_p}\left(Z_pc_F^{(\mu)}{\bf L\cdot S}_\mu+Z_{\mu}c_F^{(p)}{\bf L\cdot S}_p\right)
\left(\frac{ e^{-\lambda r}(1+\lambda r)}{r^3}\right)\\
&&
\nn
\left.
+
\frac{\alpha c_F^{(\mu)}c_F^{(p)}}{m_\mu m_p}
\left(-\frac{2}{3}\frac{ e^{- \lambda r } \lambda ^2}{ r}{\bf S_\mu\cdot S_p}+\frac{8}{3} \pi \delta^{(3)}(\vec r){\bf S_\mu\cdot S_p} +\frac{e^{- \lambda r} }{4 r^3}\left(1+ r \lambda +\frac{r^2 \lambda ^2}{3}\right)\hat{S}_{p\mu}(\hat{\bf r})\right)
\right\},
\eea
where $\lambda=m_e q$. 
Therefore, with the precision we aim at, we obtain
\be
V^{(2)}=V_{\rm tree}^{(2)}+V^{(2,2)}_{\rm 1-loop}+(V^{(2)}_{\rm VP,1-loop}+V^{(2,2)}_{\rm off-shell})+{\mathcal O}(V^{(2,3)})
=V^{(2,1)}+V^{(2,2)}+{\mathcal O}(V^{(2,3)}),
\ee
where in the second equality we organize the potential terms according to their powers in $\al$. This requires expanding the NRQCD Wilson coefficients in powers of $\al$. The leading non-vanishing contribution reads\footnote{Strictly speaking there could still be some ${\mathcal O}(\al)$ included in $\kappa_p^{\rm had}$ 
with the definition we are using, similarly 
as it happens with the proton radius.}
\begin{eqnarray}
\label{V21}
 V^{(2,1)}&=&\frac{Z\alpha }{2 {m_\mu m_p}}
 \left[-\left\{\frac{1}{ r},\bf{p}^2\right\}+\frac{1}{ r^3}{\bf L}^2+4\pi \left(1+\frac{{m_\mu^2+m_p^2}}{{4m_\mu m_p}}\right)\delta^{(3)}({\bf r})\right]
\\
&& 
+ \frac{\alpha }{2 {m_\mu}m_p}
\left[
 \frac{16}{3}\pi Z_{\mu}(Z_p+\kappa_p^{\rm had}){\bf S_\mu S_p}
 \delta^{(3)}({\bf r})
 +Z_{\mu}\frac{Z_p+\kappa_p^{\rm had}}{2}\frac{1}{ r^3} \hat{S}_{p\mu}(\hat{\bf r})
 \right]
 \nonumber 
\\
\nn
&+&
\frac{\alpha }{2 {m_\mu m_p}}
\left[
Z_{\mu}\left(2(Z_p+\kappa_p^{\rm had})+\frac{m_\mu}{ m_p}
(Z_p+2\kappa_p^{\rm had})\right)\frac{1}{ r^3}{\bf L\cdot S}_p+Z\left(2+\frac{m_p}{m_\mu}\right)\frac{1}{ r^3}{\bf L\cdot S}_\mu
\right]
\\
\nn
&+&
 \frac{\pi\al }{2m_p^2}Z_{\mu}\left[\frac{4}{3} r_p^2 m_p^2\right]\delta^{(3)}({\bf r})
.
\end{eqnarray}

For the organization of the computation it is also convenient to split $V^{(2,2)}$ in the following way:
\be
V^{(2,2)}=V^{(2,2)}_{\rm no-VP}+V^{(2,2)}_{\rm VP}.
\ee 
The first term is the potential if we switch off the interaction with the electrons. This is a well defined limit, as it corresponds to the case of muonium. 
The second term is the correction to the potential associated to the one-loop electron vacuum polarization. 

Finally, the $1/m^3$ potential, which we directly consider in position space, just comes from the Taylor expansion of the relativistic expression of the dispersion relation:
\begin{equation}
\label{V30}
V^{(3,0)}=-\frac{1}{8}\left(\frac{1}{ m_\mu^3}+\frac{1}{ m_p^3}\right){\bf p}^4
\,.
\end{equation}
There are no ${\mathcal O}(\al/m^3)$ terms.

\section{Muonic hydrogen Lamb shift: \texorpdfstring{$E(2P_{3/2})-E(2S_{1/2})$}{}}
\label{chapter3}

In this section we review all (and check several of) the contributions to the energy shift of order 
$\al^5$, as well as those that scales like $\al^6\times$logarithms in the 
context of pNRQED. 
The muonic hydrogen Lamb shift is obtained by 
the combined use of non-relativistic quantum mechanics perturbation theory and perturbative quantum field theory computations (when ultrasoft photons 
show up).  
As we have definite counting rules to asses the relative importance of the different terms we know when we can stop computing. 
The application of this program to the muonic hydrogen produces the contributions we use in our analysis, listed in Table \ref{table}. Most of the results were already available in the literature, we have checked many. 
We now discuss them focusing on the novel aspects. 
Note that, even though most of the contributions can be 
associated to a pure QED calculation, the hadronic effects are also included in this computation. Their effects are included in the NRQED Wilson coefficients discussed in Sec.~\ref{NRQED(mu)}, and are encoded in the different terms of the potential in the Lagrangian of pNRQED discussed in Sec.~\ref{Sec:pNRQCD}.  

In order to carry out the computations of this paper we use the most updated PDG values \cite{Agashe:2014kda} for the masses and fine structure constant 
\begin{eqnarray}
m_e&=&
   0.510998928(11) \,{\rm MeV},\nonumber\\
m_\mu&=&
   105.6583715(35) \,{\rm MeV}
,\nonumber\\
\alpha &=&1/137.035 999 074(44),\nonumber\\
m_p&=&
   938.272046(21)  \,{\rm MeV}
,\nonumber\\
m_\tau&=&
  1776.82(16)  \,{\rm MeV}
.
\label{masses}
\end{eqnarray}
These numbers update the values used in Ref.~\cite{Peset:2014yha}. The effect is very small but it changes the last digit of our numbers in some cases after rounding. This happens in the second term in 
Eqs.~(\ref{El1}-\ref{El2}), in Eq.~(\ref{rpfinal}), and in the i) and vii) entries in the table \ref{table}. 

$\rho_{nl}$ is the non-relativistic charge density of the $n l$ state. For the $n=2$ Lamb shift we will need their difference between the  
$S$- and $P$-wave bound state: 
\begin{eqnarray}
\rho&\equiv&\rho_{2P}-\rho_{2S}=(m_r  Z\alpha )^{3/2} e^{-m_r  Z\alpha  r} \left[\frac{1}{12} (m_r  Z\alpha  r)^2-\left(1-\frac{m_r  Z\alpha  r}{2}\right)^2\right].
\end{eqnarray}

We will use the following notation: 
\be
\delta E_{nlj}^{V}=\langle nlj | V | nlj \rangle
\ee
and
\be
\delta E_L^{V}=\langle 2P_{1/2} | V | 2P_{1/2} \rangle -
\langle 2S_{1/2} | V | 2S_{1/2} \rangle
\ee
for the correction to the Lamb shift of a generic potential $V$.

\begin{figure}[ht]
  \begin{center}
   \includegraphics[width=0.45\textwidth]{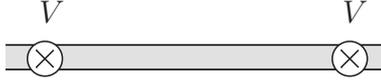}
  \end{center}
  \caption{\it 2nd order perturbation theory of the bound-state Green function generated by a generic potential $V$.}
  \label{fig:doubleVP}
\end{figure}

We will represent the 2nd order perturbation theory correction to the bound-state Green function generated by a generic potential $V$ by Fig~\ref{fig:doubleVP}, 
where the double line represents the bound state and the vertices (local in time) 
the potentials. In case we 
want to obtain the associated energy shift we will compute objects like (and analogous expressions in case of different potentials (including permutations))
\bea
\nn
\delta E_{nlj}^{V\times V}&=&
\langle \psi_{nlj}|V\frac{1}{{(E_{n}-h)'}}V|\psi_{nlj}\rangle
\\
&=&
\int \mathrm{d}{\bf r_2} \mathrm{d}{\bf r_1} \psi_{nlj}^*({\bf r_2})V({\bf r_2})
G'_{nl}({\bf r_1},{\bf r_2})V({\bf r_1})\psi_{nlj}({\bf r_1})
\label{nloV},
\eea
where 
\be
\frac{1}{ (E_{n}-h)'}=\lim_{E \rightarrow E_n}\left(\frac{1}{E-h}-\frac{1}{E-E_n}\right)
\,,
\ee
\begin{equation}
G'_{nl}({\bf r_1},{\bf r_2})
\equiv 
\langle {\bf r}_1| \frac{1}{{(E_{nl}-h)'}}|{\bf r}_2 \rangle
\equiv
\lim_{E\rightarrow E_n}\left(G({\bf r_1},{\bf r_2}; E)-\frac{|\psi_{nl} |^2}{E-E_n}\right),
\end{equation}
$\psi_{nl}({\bf r})$ is the bound state wave function of the ($n l$)-state and $E_n$ is the energy of the state, 
and $G({\bf r_1},{\bf r_2}; E)$ is the Coulomb Green function. 

In order to perform the computation it is specially useful to use the following representation for a negative energy $E=-\frac{{m_r Z^2\alpha^2}}{{2\lambda^2}}$ of the Coulomb Green function (see for instance, the appendix of Ref.~\cite{Pineda:2011dg}):
\begin{eqnarray}
G({\bf r_1},{\bf r_2}; E)&=&\frac{{m_r ^2Z\alpha }}{{\lambda\pi}}
\sum_{l=0}^{\infty}(2l+1)P_l(\frac{{{\bf r_1}\cdot{\bf r_2}}}{{r_1 r_2}})(\frac{{2m_rZ\alpha}}{{\lambda}}r_1)^l(\frac{{2m_rZ\alpha}}{{\lambda}}r_2)^le^{-\frac{{m_rZ\alpha}}{{\lambda}}(r_1+r_2)}
\nonumber \\
&&\sum_{s=0}^\infty \frac{{L_s^{2l+1}(\frac{{2m_rZ\alpha}}{{\lambda}}r_1)L_s^{2l+1}(\frac{{2m_rZ\alpha}}{{\lambda}}r_2)s!}}{{(s+l+1-\lambda)(s+2l+1)!}}\label{G}.
\end{eqnarray}
Then $G'_{nl}({\bf r_1},{\bf r_2})$ is just the Coulomb Green function evaluated at $\lambda=n+\delta\lambda$, and taking out the pole. In the case where the potentials that appear in Eq.~(\ref{nloV}) are only functions of the modulus of ${\bf r}$ (i.e. they are rotational 
invariant), the sum over $l$ reduces to the single term that matches the angular
momentum $l$ of the bound state.  

Obviously a similar discussion applies to higher order corrections from perturbation theory, and also similar expressions follow for the Lamb shift. 

We will now study each relevant contributing term separately, both in the $1/m_\mu$ and in the $\alpha$ expansions. We will write explicitly the 
$Z_{\mu}$, $Z_p$, $Z$ dependence except for the dependence on $Z$ 
that appears in the combination $m_rZ\al/m_e$ in the numerical integrals we 
perform. Therefore, such numerical values will change for different muonic atoms.

\subsection{Corrections from the static potential: \texorpdfstring{$V^{(0)}$}{}}

\subsubsection{One-loop Vacuum Polarization: \texorpdfstring{$\delta E_L^{V_{\rm VP}^{(0,2)}} \sim \mathcal O (m_\mu\alpha^3)$}{}}

The Lamb shift in muonic hydrogen, unlike in hydrogen, receives its most important contribution from the electron vacuum polarization. This is due to the fact that the typical atomic momentum of the muonic hydrogen is $m_\mu \alpha$, which is of the order of the electron mass: $m_\mu \alpha\sim 1.5 m_e$. This effect comes from the modification of the photon propagator, as we have already seen in the previous chapter (see Fig. \ref{fig:VP}). In order to compute it, we must take the first order in $\alpha$ of the expansion of $\Pi(-{\bf k}^2)$.

The integral in $r$ and $x$ can be done analytically. The result reads (see for instance \cite{Jentschura:2011nx})
\begin{align}
\label{Evp}
& \delta E_L^{V_{\rm VP}^{(0,2)}} =\int d^3r V^{(0,2)}_{\rm VP}(r) (\rho_{2P}-\rho_{2S})=
\\[2ex]
& = \frac{\alpha}{\pi} \, (Z\alpha)^2 \, m_r \left[
\frac{8 \pi \beta^3}{3} 
+ \frac{1 - 26 \beta^2 + 352 \beta^4 - 768 \beta^6}%
{18 \, (1 - 4 \, \beta^2)^2} \right.
\nonumber\\
\nn
& \; \left. + \frac{4 \beta^4 \left( 15 - 80\beta^2 + 128\beta^4 \right) }%
{3 \, (1 - 4 \, \beta^2)^{5/2}} \,
\ln\left( \frac{1- \sqrt{1- 4\beta^2}}{2 \beta} \right) \right]=m_r\alpha^3Z^20.005555
=205.00737\, \mathrm{meV},
\end{align}
where
\begin{equation}
\beta = \frac{m_e}{(Z \alpha \, m_r)}= 0.7373836 \,.
\end{equation}
Eq.~(\ref{Evp}) gives the first entry in Table~\ref{table}.

For the case $m_e \ll Z \alpha \, m_r$ the computation can be checked with the result of heavy quarkonium. We have checked it. 
We also observe that $m_e \ll Z \alpha \, m_r$ is a bad approximation to this quantity, so we will not consider it further but only for checking.
Actually, neither the $\beta \ll 1$ nor the $\beta \gg 1$ give a good approximation to Eq.~(\ref{Evp}).
\subsubsection{Two-loop Vacuum Polarization: \texorpdfstring{$\delta E_L^{V_{\rm VP}^{(0,3)}} \sim \mathcal  O(m_r\alpha^4)$}{} }\label{2loop}

We now compute the $\mathcal O(m_r\alpha^4)$ contribution associated to the two-loop 
static potential. We obtain the second entry of Table~\ref{table}:
\be
\label{V02}
\delta E_L^{V_{\rm VP}^{(0,3)}}=m_r\alpha^4Z^2\,
0.005599695
=1.50795\, \mathrm{meV}\,.
\ee
It agrees with the result of Pachucki \cite{Pachucki96} with 5 significant digits.

We observe that this contribution is significantly bigger than the one coming from double insertions of the leading vacuum polarization potential discussed 
in the next section. In a different context a similar situation has been found in heavy quarkonium physics \cite{Hoang:2000fm}.

\subsubsection{ Double Vacuum Polarization: \texorpdfstring{$\delta E_L^{V_{\rm VP}^{(0,2)}\times V_{\rm VP}^{(0,2)}} \sim \mathcal O(m_r\alpha^4)$}{}}

The second ${\mathcal O}(m_r\alpha^4)$ correction is generated by the second order perturbation theory of the $V_{\rm VP}^{(0,2)}$ potential. Following 
Eq.~(\ref{nloV}) and the associated discussion
we obtain
\begin{eqnarray}
&&\delta E_{nl}^{V_{\rm VP}^{(0,2)}\times V_{\rm VP}^{(0,2)}}=\langle \psi_{nl}|V^{(0,2)}_{\rm VP}\frac{1}{{(E_{nl}-h)'}}V^{(0,2)}_{\rm VP}|\psi_{nl}\rangle  \nonumber\\
&&=
(Z\alpha)^2 m_r 
\left(\frac{\alpha}{\pi}\right)^2\int_4^\infty \int_4^\infty d\rho_1^2 d\rho_2^2\frac{u\left(\rho_1^2\right)}{\rho_1^2}\frac{u\left(\rho_2^2\right)}{\rho_2^2}  
\\
\nn
&&
\times
\left(\frac{[1+\beta(\rho_1+\rho_2)]^{ -3}{\beta^2}Q}{12 (1+\beta \rho_1)^5(1+\beta \rho_2)^5}+\frac{\beta^2[\rho_1^2+\rho_2^2+{2}(\beta\rho_1\rho_2)^2]}{(1+\beta \rho_1)^4(1+\beta \rho_2)^4}\ln\left(\frac{(1+\beta \rho_1)(1+\beta\rho_2)}{1+\beta (\rho_1+\rho_2)}\right)\right),
\nn
\eea
where
\bea
Q&=&12 \beta ^8 \rho_1^6 \rho_2^4{+}12 \beta ^8 \rho_1^5 \rho_2^5+36 \beta ^7 \rho_1^6 \rho_2^3+120 \beta ^7 \rho_1^5 \rho_2^4+12 \beta ^6 \rho_1^6 \rho_2^2+84 \beta ^6 \rho_1^5 \rho_2^3+74 \beta ^6 \rho_1^4 \rho_2^4\nn\\
&+&33 \beta ^5 \rho_1^6 \rho_2+39 \beta ^5 \rho_1^5 \rho_2^2-62 \beta ^5 \rho_1^4 \rho_2^3+9 \beta ^4 \rho_1^6+111 \beta ^4 \rho_1^5 \rho_2-33 \beta ^4 \rho_1^4 \rho_2^2-142 \beta ^4 \rho_1^3 \rho_2^3\nn\\
&+&24 \beta ^3 \rho_1^5+99 \beta ^3 \rho_1^4 \rho_2-189 \beta ^3 \rho_1^3 \rho_2^2+18 \beta ^2 \rho_1^4-3 \beta ^2 \rho_1^3 \rho_2-75 \beta ^2 \rho_1^2 \rho_2^2-24 \beta  \rho_1^2 \rho_2-3 \rho_1^2\nn\\
&+&(\rho_1\rightarrow \rho_2).
\eea
This expression corrects several mistakes in Eq. (11) of \cite{Jentschura:2011ck} (which however gets the correct numerical result) and yields the 3rd entry 
of Table~\ref{table}:
\be
\label{V0V0}
\delta E_L^{V_{\rm VP}^{(0,3)}\times V_{\rm VP}^{(0,3)}}
=m_r\alpha^2Z^2\left(\frac{\alpha}{\pi}\right)^2 0.0055304=0.150897\; \text{meV}.
\ee
This numerical value agrees with \cite{Pachucki96} within the significant digits given in this reference.

\subsubsection{Static potential (vacuum polarization): $\delta E_L \sim$ 
\texorpdfstring{$\mathcal O(m_r\alpha^5)$}{}}

The first four entries in Table~\ref{table} are the contributions to the Lamb shift associated to the electron vacuum polarization corrections to the static potential $V^{(0)}$. Specially difficult is the 4th entry, as it corresponds to the three-loop static potential and to the third order computation in perturbation theory. It was computed (numerically) 
in \cite{Kinoshita:1998jf} (see also \cite{Ivanov:2009aa} for a small correction). It can be split into the following contributions:
\bea
\delta E_L^{V^{(0,4)}_{\rm VP}}
=m_r\alpha^5Z^20.002694 =\, 5.295 \times 10^{-3}\mathrm{meV}.
\label{V04VP}
\eea

The contribution from 2nd order perturbation theory yields \cite{Kinoshita:1998jf} (this result includes all permutations):
\bea
\delta E_L^{V_{\rm VP}^{(0,2)}\times V_{\rm VP}^{(0,3)}}+\delta E_L^{V_{\rm VP}^{(0,3)}\times V_{\rm VP}^{(0,2)}}
=
m_r\alpha^5Z^2 0.00109562 =\, 2.153 \times 10^{-3}\mathrm{meV}.
\label{V02V03VP2rd}
\eea

And the contribution from 3rd order perturbation theory reads \cite{Ivanov:2009aa,Kinoshita:1998jf}:
\bea
\delta E_L^{V_{\rm VP}^{(0,2)}\times V_{\rm VP}^{(0,2)}\times V_{\rm VP}^{(0,2)}}
=
m_r\alpha^5Z^2 0.0000377=\, 0.0741 \times 10^{-3}\mathrm{meV}.
\label{V02V02V02VP3rd}
\eea

The sum of the above three terms gives the final contribution:
\begin{equation}
\label{V03}
\delta E_{L,\rm static}^{{\mathcal O}(\al^5),{\rm VP}}=0.11868m_r Z^2\alpha^2 \left( \frac{\alpha}{\pi}\right)^3=0.00752\, \mathrm{meV}.
\end{equation}
which is the 4th entry of Table~\ref{table}. 
The computation has been done independently for a time-like ($q^2>0$) and a space-like ($q^2<0$) momentum of the photon. This last one involves the integration of the whole vacuum polarization function $\Pi (q^2)$ to the desired order, and the other involves just its imaginary part evaluated at $q^2=t\, m_e^2$. 

\subsubsection{Static potential (light-by-light): \texorpdfstring{$\delta E_L^{V_{\rm LbL}^{(0,4)}}
 \sim \mathcal O(m_r\alpha^5)$}{} }
 
The 5th entry of Table \ref{table} corresponds to the contribution associated to the light-by-light corrections to the static potential $V^{(0)}$, i.e. to 
$V_{\rm LbL}$ (see Eq. (\ref{VLbL})). It was obtained in \cite{Karshenboim:2010cq}, where a very long explanation was made to argue that the light-by-light
contributions could be computed in the static approximation. This is evident in the EFT, as they correspond to a correction to the static potential, as already stated in Ref.~\cite{Pineda:2004mx}.

The result for this contribution, given in \cite{Karshenboim:2010cq}, is
\bea
\nn
\delta E_L^{V^{(0,4)}_{\rm LbL} }
&=&-m_r\alpha^5Z^2 \,10^{-3}\left[(Z_p^2+Z_{\mu}^2)0.5185 -Z_pZ_{\mu} 0.5852\right]
\\&=&-m_r\alpha^5\, 0.000452 =-0.00089(2)\, \mathrm{meV}.
\label{V0LbL}
\eea

\subsection{Corrections from the \texorpdfstring{$1/m_\mu$}{} potentials without vacuum polarization}

We jump directly into the $V^{(2)}$ potential, since we already discussed that the $V^{(1)}$ potential produces corrections of, utmost, $\mathcal O (m_r\alpha^6)$ and are then beyond the accuracy of our interest.

We now compute the corrections to the energy and Lamb shift associated to
the potentials in Eqs.~(\ref{V2}) and 
(\ref{V221loop}) to ${\mathcal O}(m_r\al^5)$. In other words, we compute the relativistic corrections that are not associated to the vacuum polarization.

\subsubsection{Relativistic corrections:  \texorpdfstring{$\delta E_L \sim \mathcal O(m_r\alpha^4)$}{} }

Eq.~(\ref{V2}) is the EFT generalization of the Breit potential. Note that 
it is in this potential where the hadronic corrections arise at $\mathcal O (m_r \alpha^5 \frac{m_\mu^2}{m_p^2})$ (we will consider them in more detail later).
The energy shift associated to this potential reads\footnote{In the last line of this equation we have still included the contribution associated to the 
tau vacuum polarization. As its numerical effect is very small we will neglect it in the following.}
\begin{eqnarray}
\delta E^{V^{(2)}_{\rm tree}}_{nljj_{\mu}}&=& 
\frac{m_r^3Z^3\alpha^4}{ 2 n^3 m_\mu^2} 
\left\{Z_p c_D^{(\mu)}\delta_{l0}+
Z_pc_S^{(\mu)}\frac{(1-\delta_{l0})}{l(l+1)(2l+1)}d_{j_\mu,l}
\right.\nonumber\\
&&
+2\frac{m_\mu}{m_p}\left(
Z\left(\frac{1}{n}+\frac{(1+4l)\delta_{l0}-3}{2l+1}\right)
+
c_F^{(\mu)}c_F^{(p)}\left(\frac{(1-\delta_{l0})\delta_{s1}}{2l(l+1)(2l+1)}c_{j,l}-2\delta_{l0}+\frac{8}{3}\delta_{l0}\delta_{s1}\right)\right.
\nonumber\\
&+&\left.\left.\frac{(1-\delta_{l0})}{l(l+1)(2l+1)}
\left(Z_p c_F^{(\mu)}d_{j_\mu,l}+Z_{\mu}c_F^{(p)}(2h_{j,l}\delta_{s1}-d_{j_\mu,l})\right)\right)\right.\nonumber\\
&+&\left.\frac{m_\mu^2}{m_p^2}\left(
Z_{\mu}c_D^{(p)}\delta_{l0}+Z_{\mu} c_S^{(p)}\frac{(1-\delta_{l0})}{l(l+1)(2l+1)}\left(2\delta_{s1}h_{j,l}-d_{j_\mu,l}\right)\right)\right\}\nonumber\\
&-&\frac{m_r^3Z^3\alpha^3}{ \pi n^3 }\delta_{l0}\left\{\frac{1}{m_p^2}\left(c_3+(3-4\delta_{s1})c_4\right)+16\pi Z\alpha\left(\frac{d_2}{m_p^2}+\frac{d_2^{(\mu)}}{m_{\mu}^2}+\frac{d_2^{(\tau)}}{m_\tau^2}\right)\right\},
\label{Etree}
\end{eqnarray}
where
\begin{eqnarray}
c_{j,l}&=&2\begin{cases}\displaystyle -\frac{l+1}{2l-1} &\text{$j=l-1$},\\1&\text{$j=l$},\\-\frac{l}{2l+3}&\text{$j=l+1$}\,, \end{cases}\label{c}\\\nonumber\\
h_{j,l}&=&\begin{cases}\displaystyle -(l+1) &\text{$j=l-1$},\\-1&\text{$j=l$},\\l&\text{$j=l+1$}\,, \end{cases}\\\nonumber\\
d_{j_1,l}&=&\begin{cases}\displaystyle -(l+1) &\text{$j_1=l-\frac{1}{2}$},\\l &\text{$j_1=l+\frac{1}{2}$}\,.
\label{d}\end{cases}
\end{eqnarray} 
The energy has been expressed in terms of the total angular momentum 
$\bf J=L+S$ (where $\bf{S}=\bf{S}_{\mu}+\bf{S}_p$) and in terms of the angular momentum of the muon 
$\bf J_\mu$=$\bf L+S_\mu$. The basis is taken in terms of the lightest particle, since it is the most convenient one to express the energy shift. This is so since the lightest particle gives rise to larger effects in the terms which involve the ratio of the masses, and this comes out more clearly when using this basis.

$\delta E^{V^{(2)}_{\rm tree}}_{nljj_{\mu}}$ encodes all the $\mathcal O(m_r\alpha^4)$ corrections to the spectrum due to the $1/m_{\mu}^2$. It also includes higher order effects through the 
${\mathcal O}(\al)$ terms in the NRQCD Wilson coefficients. If we set them to zero, we obtain the non-trivial leading-order contribution:
\begin{eqnarray}
\delta E^{V^{(2)}_{\rm tree}}_{nljj_{\mu}}&=& 
\frac{m_r^3Z^3\alpha^4}{ 2 n^3 m_\mu^2} 
\left\{Z\delta_{l0}+
Z\frac{(1-\delta_{l0})}{l(l+1)(2l+1)}d_{j_\mu,l}
\right.\nonumber\\
&&
+2\frac{m_\mu}{m_p}\left(
Z\left(\frac{1}{n}+\frac{(1+4l)\delta_{l0}-3}{2l+1}\right)
\right.
\nonumber\\
&&
+
Z_{\mu}(Z_p+\kappa_p^{\rm had})\left(\frac{(1-\delta_{l0})\delta_{s1}}{2l(l+1)(2l+1)}c_{j,l}-2\delta_{l0}+\frac{8}{3}\delta_{l0}\delta_{s1}\right)
\nn
\\
&&+\left.\left.\frac{(1-\delta_{l0})}{l(l+1)(2l+1)}
\left(Z d_{j_\mu,l}+Z_{\mu}(Z_p+\kappa_p^{\rm had})
(2h_{j,l}\delta_{s1}-d_{j_\mu,l})\right)\right)\right.\nonumber\\
&&+\left.\frac{m_\mu^2}{m_p^2}\left(
Z\delta_{l0}+Z_{\mu} (Z_p+2\kappa_p^{\rm had})\frac{(1-\delta_{l0})}{l(l+1)(2l+1)}\left(2\delta_{s1}h_{j,l}-d_{j_\mu,l}\right)\right)\right\}\nonumber
\\
&&+\left[\frac{4}{3} r_p^2  m_p^2\right]\frac{\pi\alpha}{2m_p^2}Z_{\mu}\delta_{l0}\frac{1}{\pi}\left(\frac{m_r Z\alpha}{n}\right)^3
\,.
\label{EV21}
\end{eqnarray}

We shall also take into account the correction to this order in $\alpha$ coming from the perturbative expansion of the relativistic kinetic term, i.e. from Eq.~(\ref{V30}), which leads to the energy shift:
\begin{eqnarray}
\label{EV30}
\delta E^{V^{(3,0)}}_{nl} &=&
{m_r^3Z^4\alpha^4\over {2m_\mu^2}}\left[\left(1-{m_\mu\over { m_p}}+\left({m_\mu\over m_p}\right)^2\right)\left({3\over {4n^4}}-{2\over{n^3(2l+1)}}\right)\right].
\end{eqnarray}

Summing up the contributions of Eqs.~(\ref{EV21}), (\ref{EV30}) we get for the transition of the $2S^{1/2}\rightarrow 2P^{1/2}$ states:
\begin{equation}
\delta E^{V^{(2,1)}}_L +\delta E^{V^{(3,0)}}_L={{m_r^3\alpha^4Z^4}\over{48m_p^2}}
-
{{m_r^3\alpha^4Z^3Z_\mu}\over{16m_p^2}}\left[\frac{4}{3} r_p^2 m_p^2\right]=\left(0.05747-5.1975\frac{ r_p^2 }{\rm fm^2}\right)\, \mathrm{meV}.
\label{Ek}
\end{equation}

The first term agrees both analytically and numerically with the one obtained in \cite{Pachucki96}. We shall remark that it 
has an extra $\frac{m_r^2}{m_p^2}$ suppression factor, which was to be expected since this correction does not contribute for the case of the hydrogen (in the infinite proton mass limit).
The 2nd term is the leading contribution associated to the proton radius. Both contributions appear as the 6th and 11th entries in Table~\ref{table}.

\subsubsection{Relativistic corrections:  \texorpdfstring{$\delta E_L \sim \mathcal O(m_r\alpha^5)$}{} }

We now compute the $\mathcal O(m_r\alpha^5)$ contributions to the spectrum with no electron vacuum polarization. As we have already mentioned, this is a well defined quantity, as it amounts to the corresponding evaluation of the muonium ($\mu e$) spectrum (if we turn off the hadronic effects). Taking the ${\mathcal O}(m_r\al^5)$ corrections generated from Eq.~(\ref{Etree}) (typically generated by the ${\mathcal O}(\al)$ corrections of the NRQCD Wilson coefficients) plus the energy shift produced by the expectation value of Eq.~(\ref{V221loop}), 
we obtain (note that this computation has been done in the $\MS$ scheme)
\begin{eqnarray}
\delta E_{nljj_{\mu}}^{V^{(2,2)}_{\rm no-VP}}&=&\frac{m_r^3Z^3\alpha^5}{2\pi n^3}\left\{\frac{Z Z_\mu^2}{m_\mu^2}\left(\frac{4}{3}\left(-\frac{2}{5}+\ln\left(\frac{m_\mu^2}{\nu^2}\right)\right)\delta_{l,0}+\frac{1-\delta_{l,0}}{l(l+1)(2l+1)}d_{j_\mu,l}\right)\right.\nn\\
&+&\left.\frac{Z_p^2Z}{m_p^2}\left(\frac{4}{3}\left(-\frac{2}{5}+\ln\left(\frac{m_p^2}{\nu^2}\right)\right)\delta_{l,0}+\frac{1-\delta_{l,0}}{l(l+1)(2l+1)}(2\delta_{s,1}h_{j,l}-d_{j_\mu,l})\right)\right.\nn\\
&+&\left.\frac{1}{m_\mu m_p}\left(-2\left( Z Z_\mu^2+ Z Z_p^2-\frac{ Z^2}{3}+Z_\mu^3 \kappa_p^{\rm had}\right)\delta_{l,0}-\frac{14}{3}Z^2\frac{1-\delta_{l,0}}{l(l+1)(2l+1)}\right.\right.\nn\\
&+&\left.\left.\frac{14}{3}Z^2\delta_{l,0}\left(1-\frac{1}{n}+2k(n)+2\ln\left(\frac{2\alpha m_r}{n \nu}\right)\right)\right.\right.\nn\\
&+&\left.\left.\frac{8}{3}\left( Z Z_\mu^2+ Z Z_p^2+Z_\mu^3 \kappa_p^{\rm had}\right)\delta_{s,1}\delta_{l,0}+\frac{1-\delta_{l,0}}{l(l+1)(2l+1)}\left(\frac{1}{2}\left( Z Z_\mu^2+ Z Z_p^2+Z_\mu^3 \kappa_p^{\rm had}\right)\delta_{s,1}c_{j,l}\right.\right.\right.\nn\\
&+&\left.\left.\left.2Z Z_p^2 \delta_{s,1}h_{j,l}+Z(Z_\mu^2-Z_p^2)d_{j_\mu,l}\frac{}{}\right)\right)+\frac{2Z^2\delta_{l,0}}{m_\mu^2-m_p^2}\left(\frac{m_p}{m_\mu}\left(\frac{1}{3}+\ln\left(\frac{m_\mu^2}{\nu^2}\right)\right)\right.\right.\nn\\
&+&\left.\left.\frac{m_\mu}{m_p}\left(\frac{1}{3}+\ln\left(\frac{m_p^2}{\nu^2}\right)\right)+\ln\left(\frac{m_\mu^2}{m_p^2}\right)(3-4\delta_{s,1})\right)-\frac{8Z_p Z_\mu }{15 m_\tau^2}\delta_{l,0}\right\}
\nn
\\
&-&\frac{m_r^3Z^3\alpha^3}{\pi n^3}\delta_{l 0}\left[\frac{1}{m_p^2}\left(c^{\rm had}_3+(3-4\delta_{s, 1}) c^{\rm had}_4\right)+16\pi\alpha \frac{d^{\rm had}_2}{m_p^2} \right]
\label{V22noVP}
,
\end{eqnarray}
where $k(n)=\sum_{k=1}^n\frac{1}{k}$ is the n-th harmonic number.
Note that in this expression the hadronic corrections that scale as $\al^2$: $c_3^{\rm had}$, $c_4^{\rm had}$ 
and $\al d_2^{\rm had}$ are also included, as they also produce an $m_r\al^5$ energy shift. 

\subsection{Ultrasoft effects: \texorpdfstring{$\delta E_L \sim \mathcal O( m_r\alpha^5)$}{} }

\begin{figure}[h]
  \begin{center}
    \includegraphics[width=0.5\textwidth]{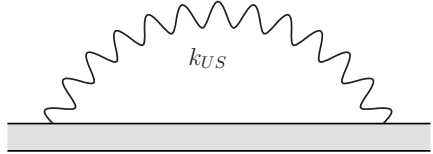}
  \end{center}
  \caption{\it Correction due to ultrasoft photons.}
  \label{fig:ultrasoft}
\end{figure}

The interaction of the bound state with ultrasoft photons (symbolically pictured in Fig.~\ref{fig:ultrasoft}) produces an energy shift of ${\mathcal O}(m_r\alpha^5)$. It has been computed in the $\MS$ in Refs.~\cite{Pineda:1997ie,Pineda:1998kn} for the case of hydrogen and positronium respectively. The application 
to muonic hydrogen is trivial, as we only have to rescale for the reduced mass. 
On top of that we introduce the changes for the case of particles with 
general charges $Z_{\mu}$, $Z_p$. Finally, the energy shift reads (in the $\MS$ scheme)
\begin{eqnarray}
\delta E^{\rm US}_{nl}&=&\frac{2}{3}
\left(\frac{Z_{\mu} m_p+Z_p m_{\mu}}{m_p+m_{\mu}}\right)^2
\frac{\alpha}{\pi}\left(\left(\ln \frac{\nu}{m_r}+\frac{5}{6}-\ln 2\right)\left(\frac{Ze^2}{2}\right)\frac{|\phi_n(0)|^2}{m_r^2}\right.\nonumber\\
&-&\left.\sum_{n'\neq n}|\langle n|\frac{p}{m_r}|n'\rangle|^2(E_n-E_{n'})\ln\frac{m_r}{|E_n-E_{n'}|}\right)
\nn
\\
&\equiv&\frac{ m_rZ^4\alpha ^5}{n^3\pi } 
\left(\frac{Z_{\mu} m_p+Z_p m_{\mu}}{m_p+m_{\mu}}\right)^2
\left(\delta_{l,0}\left(-\frac{4}{3}\left(\ln R(n,l)+\ln\frac{m_rZ^2\al^2}{\nu}
\right)+\frac{10}{9}\right)\right.\nonumber\\
&-&\left.(1-\delta_{l,0})\frac{4}{3}\ln R(n,l)\right),
\label{Eus1}\end{eqnarray}
where $|\phi_n(0)|^2=\frac{1}{\pi}\left(\frac{m_rZ\al}{n}\right)^3$. $\ln R(n,l)$ are the Bethe logarithms and are implicitly defined by the equality with Eq.~(\ref{Eus1}). For their numerical values for the $2S$ and $2P$ states we have used the values quoted in \cite{Pachucki96}.

We observe that $\delta E^{\rm US}_{n,l}$ is factorization scale dependent. Such dependence cancels with the 
scale dependence of Eq.~(\ref{V22noVP}). The sum of both contributions gives all the ${\mathcal O}(m_r\alpha^5)$
corrections to the spectrum that are not associated to the electron vacuum polarization:
\begin{equation}
\delta E^{{\mathcal O}(\al^5),{\rm no-VP}}_{nljj_{\mu}}=
\delta E_{nljj_{\mu}}^{V^{(2,2)}_{\rm no-VP}}+\delta E^{\rm US}_{nl}
\,,
\end{equation}
and is independent of the factorization scale. 
It can also be 
split into the different hadronic contributions, associated to the fact that the proton is not point-like, and  the 
${\mathcal O}(\al^5)$ contribution to the spectrum of two point-like particles (relevant for muonium) in the following way: 
\begin{equation}
\delta E^{{\mathcal O}(\al^5),{\rm no-VP}}_{nljj_{\mu}}=
\delta E^{{\mathcal O}(\al^5),{\rm no-VP}}_{nljj_{\mu},{\rm point-like}}
+\delta E^{{\mathcal O}(\al^5),{\rm no-VP}}_{nljj_{\mu},d_2^{\rm had}}
+
\delta E^{{\mathcal O}(\al^5),{\rm no-VP}}_{nljj_{\mu},c_3^{\rm had}}
+
\delta E^{{\mathcal O}(\al^5),{\rm no-VP}}_{nljj_{\mu},c_4^{\rm had}}
.
\end{equation}

Similar equations follow for the Lamb shift energy splitting: $\delta E^{{\mathcal O}(\al^5),{\rm no-VP}}_{L}$, although in this last case the contribution proportional to $c_4^{\rm had}$ vanishes, since the spin-dependent term does not contribute to the average energy over polarizations. 
 
The above computation keeps the complete proton and muon mass dependence. It is interesting to see the convergence of the $m_{\mu}/m_p$ expansion. We do so for $\delta E^{{\mathcal O}(\al^5),{\rm no-VP}}_{L,{\rm point-like}}$, which has a non-trivial dependence on this ratio. We obtain\\

- $\mathcal O (m_\mu \alpha^5):
 $ \,\,\,\,\,\,\,\,\,\,\,\,\,\,\,\, $\delta E^{{\mathcal O}(\al^5),{\rm no-VP}}_{L,{\rm point-like}}=-0.900\, \mathrm{meV}$

- $\mathcal O (m_\mu \alpha^5\frac{m_\mu}{m_p}): $ \,\,\,\,\,\,\,\, $\delta E^{{\mathcal O}(\al^5),{\rm no-VP}}_{L,{\rm point-like}}=\,\,\,\,0.226\, \mathrm{meV}$

- $\mathcal O (m_\mu \alpha^5 \frac{m_\mu^2}{m_p^2}): $ \,\,\,\,\,\,\,\, $\delta E^{{\mathcal O}(\al^5),{\rm no-VP}}_{L,{\rm point-like}}=-0.054\, \mathrm{meV}$

- $\mathcal O( m_\mu\alpha^5 \frac{m_\mu^3}{m_p^3}): $ \,\,\,\,\,\,\,\, $\delta E^{{\mathcal O}(\al^5),{\rm no-VP}}_{L,{\rm point-like}}=\,\,\,\,0.010\, \mathrm{meV}$, \\\\
which, summing up to all orders, leads to the following ${\mathcal O}(\al^5)$ energy contribution to the Lamb shift
\begin{equation}
\label{softUS}
\delta E^{{\mathcal O}(\al^5),{\rm no-VP}}_{L,{\rm point-like}}=-0.71896 \, \mathrm{meV},
\end{equation}
which corresponds to the 7th entry of Table~\ref{table}.
This result is very similar to the one computed by Pachucki \cite{Pachucki96}, where these effects  sum up to $E(2P_{1/2}-2S_{1/2})=-0.663-0.045-0.010=-0.718\,\mathrm{meV}$ at $\mathcal O (m_\mu \alpha^5)$. 

We now consider the hadronic corrections. The energy shift associated to the hadronic vacuum polarization reads
\be
\label{VPhad}
\delta E^{{\mathcal O}(\al^5),{\rm no-VP}}_{L,d_2^{\rm had}}
=
{16\al Z d_2^{\rm had}  \over m_p^2}
\left(
{m_rZ \al \over n}
\right)^3
=0.0111(2)
\, \mathrm{meV},
\ee
which corresponds to the 14th entry of Table~\ref{table}. 

The energy shift associated to $c_3^{\rm had}$ is usually named the two-photon exchange contribution. Using 
the lower value in Eq.~(\ref{eq:c3had}) we obtained 
\be
\label{ETPE}
\delta E^{{\rm TPE}}_{L}
\equiv
\delta E^{{\mathcal O}(\al^5),{\rm no-VP}}_{L,c_3^{\rm had}}
=
{c_3^{\rm had}  \over m_p^2}
\frac{1}{\pi}
\left(
{m_r Z \al \over n}
\right)^3
=0.0344(125) \, \mathrm{meV},
\ee
which corresponds to the 15th entry of Table~\ref{table}. 

\subsection{\texorpdfstring{$1/m_{\mu}^2$}{} \texorpdfstring{electron vacuum polarization corrections: 
$\delta E_L \sim \mathcal O(m_r\alpha^5)$}{} }

We now compute the energy shifts, with $\mathcal O(m_r\alpha^5)$ precision, associated to the electron vacuum 
polarization. They are 
produced by 2nd order non-relativistic quantum mechanics perturbation theory of $V^{(0,2)}_{\rm VP}\sim \al^2/r$,  
together with the $V^{(2,1)}\sim \al/m^2$ and $V^{(3,0)} \sim 1/m^3$ potentials, as well as by the correction due to the $V^{(2,2)}_{\rm VP}\sim \al^2/m^2$ potential. This sum constitutes a well defined set, as it 
can be parametrically distinguished from other contributions (formally through the number of light fermions). 
The energy shift then reads
\begin{eqnarray}
\label{EV22VP}
&&
\delta E_{nl}^{V^{(2,2)}_{\rm VP}}+\delta E_{nl}^{V^{(2,1)}\times V_{\rm VP}^{(0,2)}}+\delta E_{nl}^{V^{(3,0)}\times V_{\rm VP}^{(0,2)}}\nn
\\
&&
=\langle \psi_{nl}| V^{(2,2)}_{\rm VP}|\psi_{nl}\rangle +2\langle \psi_{nl}|(V^{(2,1)}
+ V^{(3,0)}){1\over{(E_{nl}-h)'}}V^{(0,2)}_{\rm VP}|\psi_{nl}\rangle. 
\label{dE}
\end{eqnarray}
For the Lamb shift corrections we obtain the explicit expressions
\begin{eqnarray}
\label{ELV22VP}
\delta E_L^{V^{(2,2)}_{\rm VP}}&=&(m_r  Z \alpha )^3\frac{\alpha}{8}\frac{\alpha}{\pi}\int_4^\infty dq^2 \frac{u(q^2)}{q^2}
\\
\nn
&&
\times
\left\{-\frac{1 }{2}\left(\frac{Z_p c_D^{(\mu)}}{m_\mu^2}+\frac{Z_{\mu}c_D^{(p)}}{m_p^2}\right)\frac{(1+2 \beta  q) (1+2 \beta  q (1+\beta  q))}{ (1+\beta  q)^4}+\frac{Z}{m_\mu m_p} \frac{1+2 \beta  q}{ (1+\beta  q)^2}\right.\nonumber\\
&&-\left.\frac{Z_p }{3}\left(\frac{c_S^{(\mu)}}{2m_\mu^2}+\frac{c_F^{(\mu)}}{m_\mu m_p}\right)\left(\frac{3 \beta  q+1}{(\beta  q+1)^3}\right)\right\}=
-\left(0.027714+0.0112 \frac{ r_p^2}{\text{fm}^2}\right)
\, \mathrm{meV}\nn\\
&&+ \mathcal O(\alpha^6)
\,,
\nn
\end{eqnarray}

\begin{eqnarray}
\nn
&&
\delta E_L^{V_{\rm VP}^{(0,2)}\times V^{(2,1)}}+\delta E_L^{V_{\rm VP}^{(0,2)}\times V^{(3,0)}}
=(m_r Z \alpha )^3\frac{\alpha}{2}
\frac{\alpha}{\pi}\int_4^\infty dq^2 \frac{u(q^2)}{q^2}
\\
&&
\times
\left\{\frac{m_r }{6}\left(\frac{Z}{m_\mu^3}+\frac{Z}{m_p^3}\right)\left(-\frac{4 \left(1+3 q^2 \beta ^2\right) }{(1+q \beta )^4}\ln\left(\frac{1}{1+q \beta }\right)\right.\right.\nonumber\\
&&\left.\left.+\frac{16+64 q \beta +53 q^2 \beta ^2+81 q^3 \beta ^3+24 q^4 \beta ^4}{4 (1+q \beta )^5}\right)+\frac{Z}{m_\mu m_p}\left(-\frac{1+4 q^2 \beta ^2 }{ (1+q \beta )^4}\ln\left(\frac{1}{1+q \beta }\right)\right.\right.\nonumber\\
&&\left.\left.+\frac{(3+11 q \beta ) \left(1+q^2 \beta ^2\right)}{4 (1+q \beta )^5}\right)+\frac{1}{2}\left(\frac{Z_p c_D^{(\mu)}}{m_\mu^2}+\frac{Z_{\mu}c_D^{(p)}}{m_p^2}\right)\left(-\frac{3+11 q \beta +4 q^2 \beta ^2+12 q^3 \beta ^3+4 q^4 \beta ^4}{4 (1+q \beta )^5}\right.\right.\nonumber\\
&&\left.\left.+\frac{1+2 q^2 \beta ^2}{ (1+q \beta )^4} \ln\left(\frac{1}{1+q \beta }\right)\right)+\frac{Z_p}{3}\left(\frac{c_S^{(\mu)}}{2m_\mu^2}+\frac{c_F^{(\mu)}}{m_\mu m_p}\right)\left(-\frac{3+11 q \beta +4 q^2 \beta ^2}{4(1+q \beta )^5}+\frac{\ln\left(\frac{1}{1+q \beta }\right)}{ (1+q \beta )^4}\right)\right\}\nn\\
&&=  \left(0.046473-0.016953 \frac{ r_p^2 }{\text{fm}^2}\right) \, \mathrm{meV}+{\mathcal O}(\al^6).
\label{2ndVP}
\end{eqnarray}
For this last result we have used \eq{G}. Summing up both contributions, Eqs.~(\ref{ELV22VP}) 
and (\ref{2ndVP}), 
we obtain 
\bea
&&
\delta E_L^{V^{(2,2)}_{\rm VP}}+\delta E_L^{V^{(2,1)}\times V_{\rm VP}^{(0,2)}}+\delta E_L^{V^{(3,0)}\times V_{\rm VP}^{(0,2)}}
\nn
\\
&&
=
m_r\alpha^5\,0.0095460-m_r\alpha^5\,0.01433 \frac{ r_p^2}{\text{fm}^2} =\left(0.018759-0.0282\frac{ r_p^2}{\text{fm}^2}\right)\, \mathrm{meV}\,.
\label{V2VP}
\eea
As we have already stated, this sum constitutes a well defined set, as it can be parametrically distinguished from other contributions (formally through the number of light fermions). This is also so for each individual term in the last equality in Eq.~(\ref{V2VP}). 
The first term corresponds to assuming the proton to be point-like (switching off the 
proton radius contribution) and gives the viii) entry in Table \ref{table}. This contribution was first computed in \cite{Pachucki96} 
and later corrected in \cite{Veitia,Borie:2012zz}. Nevertheless, a different number has been obtained in two recent analyses 
\cite{Jentschura:2011nx,KIK}. We confirm this last number, which is the one we quote in Table~\ref{table}. 

The term proportional to the proton radius in Eq. (\ref{V2VP}) emanates from the coefficient $c_D^{(p)}$. It 
corresponds to the xi) entry of the table and it is in agreement with the result 
found in \cite{Pachucki96}.
 
\subsection{\texorpdfstring{${\mathcal O}(m_r\al^6\times \ln)$}{} effects}

The first 8 entries in Table \ref{table} give the complete ${\mathcal O}(m_r\al^5)$ result for a point-like proton. 
In this result we have kept the exact mass dependence.
The ${\mathcal O}(m_r \alpha^6)$ contribution is dominated by the logarithmic enhanced terms. 
Here, we compute the leading ones. We assign a general counting of $ m_r/m_p \siml \ln \al \sim \ln (m_e/m_{\mu})$.  Therefore, we only compute those 
contributions at leading order in the $m_r/m_p$ expansion, i.e. those where the proton is infinitely massive. If we switch off electron vacuum polarization 
 effects (i.e. we switch off the interaction with the electron) the system corresponds to the standard hydrogen situation, which has no ${\mathcal O}(m_r\alpha^6\ln\alpha)$ effects. Actually, this is also true if we consider the case of muonium (with finite recoil effects), which again has no ${\mathcal O}(m_r\alpha^6\ln\alpha)$ effects. The reason is that the sum of all possible contributions vanishes for the case of the lamb shift, since the effective energy shift is \cite{KMY}\footnote{For simplicity we set $Z_p=Z_{\mu}=1$ in this section.}
\be
\delta E_{nls}=\frac{1}{3}\frac{m_r^5}{m_p^2m_{\mu}^2}\al^6\ln\frac{1}{\al}
\left(\delta_{s1}-\frac{3}{4}\right)
\frac{\delta_{l 0}}{n^3}
,\ee
which vanishes for the Lamb shift. 
Therefore, we can actually claim that all the ${\mathcal O}(m_r\alpha^6\ln\alpha)$ 
logarithms are generated by the electron 
vacuum polarization (for a point-like proton). Note that this would also be true if we consider proton-recoil corrections. In any case, as we have already mentioned, here we only consider the infinite proton mass limit. In this limit, for a point-like proton, only two contributions are produced 
(both of them generated by electron vacuum polarization effects), listed in the the ix) and x) entries of Table~\ref{table}, which we now discuss.

The 9th entry is due to the logarithmic-enhanced ${\mathcal O}(\al^2)$ corrections to the $c_D^{(\mu)}$ Wilson coefficient (see Eq.~(\ref{cDmu})) 
that appear in the tree-level potential (see the $c_D^{(p)}$-dependent term of Eq.~(\ref{V2})). 
It produces an $\al^3/m_{\mu}^2\times$logarithm-potential, the expectation value of which gives the following energy shift to the spectrum
\be
\delta E_{nl}=
{m_r^3\alpha^4\over {2m_\mu^2}}{c_D^{(\mu)}\over n^3}\delta_{l0}\Bigg|_{{\mathcal O}(\al^6\ln)},
\ee
and to the Lamb shift
\bea
\label{m6cDmutree}
\delta E_L&=&-m_r\alpha^6\,0.08885 =-0.0012741\,\mathrm{meV}
,
\eea
which is the number that we quote in the 9th entry of Table~\ref{table}.

The 10th entry in Table~\ref{table} is generated in the same way as the 8th entry but multiplied by 
the (logarithmic enhanced) ${\mathcal O}(\al)$ term of $c_D^{(\mu)}(\nu)$ (see Eqs.~(\ref{cDmu}) and (\ref{ELV22VP})): 
\begin{eqnarray}
\delta E_L^{V^{(2,2)}_{\rm VP}}\Bigg|_{{\mathcal O}(\al^6)}&=&-(m_r  \alpha )^3\frac{\alpha}{16}\frac{\alpha}{\pi}\int_4^\infty dq^2 \frac{u(q^2)}{q^2}
\frac{c_D^{(\mu)}}{m_{\mu}^2}\left\{\frac{(1+2 \beta  q) (1+2 \beta  q (1+\beta  q))}{ (1+\beta  q)^4}\right.\nn\\
&+&\left.\frac{4 \beta ^4 q^4+12 \beta ^3 q^3+4 \beta ^2 q^2+11 \beta  q+3}{(\beta  q+1)^5}-\frac{4 \left(2 \beta ^2 q^2+1\right) \ln \left(\frac{1}{\beta  q+1}\right)}{(\beta  q+1)^4}\right\}\Bigg|_{{\mathcal O}(\al^6\ln)}.
\end{eqnarray}
The $\nu$ dependence gets regulated by the 
ultrasoft scale, which we set to $\nu=m_{\mu}\al^2$, producing the number 
\begin{eqnarray}
\label{m6cDmuVP}
\delta E_L^{V^{(2,2)}_{\rm VP}}\Bigg|_{{\mathcal O}(\al^6\ln)}
=-m_r\alpha^6\,0.31644 =-0.004538\, \mathrm{meV},
\end{eqnarray}
which we quote in the 10th entry in Table~\ref{table}. 

Both computations were considered before in Ref. \cite{Pachucki96}. We agree with them for the significant digits given in that reference. It is also interesting to see that both contributions can be understood from a renormalization group analysis in some appropriate limit \cite{Pineda:2002bv}. This analysis also gives us 
information on the logarithmic structure of the recoil, $m_r/m_p$, corrections. At this order extra logarithmic terms appear. 
Nevertheless, they are at most linear: ${\mathcal O}(m_{\mu}\alpha^6\frac{m_r}{m_p}\ln\al)$, i.e. there are no 
${\mathcal O}(m_r\alpha^6\frac{m_{\mu}}{m_p}\ln^2\al)$ terms, contrary to the claim in Ref. \cite{Jentschura:2011nx}. 

For a point-like proton this computation would finish our analysis. The error would be due to uncomputed contributions of  ${\mathcal O}(m_r\al^6)$ and
${\mathcal O}(m_r\alpha^6\frac{m_{\mu}}{m_p}\ln\al)$. In Refs.~\cite{Jentschura:2011ck,Korzinin:2013uia} 
several terms of this order were computed. We use these analyses to estimate the error. Specially useful to us are 
the (a) and (d) entries in Table IV of the last reference. They are related with the large logarithmic contributions discussed above but also include some finite pieces. We take the difference with the pure logarithmic terms for the generic ${\mathcal O}(m_r\al^6)$ error. Taking instead 1/2 of 
the sum of the 9th and 10th entries yields a similar error: $\sim 3$ $\mu$eV. This is the error we quote in 
the first term of Eq.~(\ref{El2}), which encodes all the QED-like contributions assuming the proton to be point-like.

We now consider the ${\mathcal O} (m_r \alpha^6\ln\al)$ correction associated to the proton radius. It scales like 
${\mathcal O} (m_r \alpha^6\ln\al\times m_r^2 \, r_p^2)$ and has been computed in \cite{Friar:1978wv}. 
Such effect would be generated by the 2nd order perturbation theory of the delta potential 
(note that a similar effect would also exist in the analogous hydrogen computation). The infrared behaviour of this computation would be regulated by the inverse Bohr radius generated by the bound state dynamics, $\sim m_{\mu}\al$. The ultraviolet behaviour gets regulated by energy scales of order $m_{\mu} \sim m_{\pi}$. This 
produces the large logarithm: $\ln((m_{\mu}\al)/ m_{\mu})=\ln \al$. 
The explicit correction reads
\be
\label{r2al2}
\delta E_L
=
\frac{2\pi}{3}\left[\frac{m_r^3\al^3}{2^3\pi}\right]r_p^2\al^3\ln \al
=-0.0014
\frac{ r_p^2 }{\rm fm^2}  
\,,
\ee
 and it is listed in the 13th entry of Table \ref{table}. We use 1/2 of this result for the error of the $r_p^2 $ coefficient in Eq.~(\ref{El1}) and Eq.~(\ref{El2}).

A complete analysis of the ${\mathcal O}(m_r\alpha^6)$ effects from an EFT perspective will be discussed elsewhere. 

\section{Summary of results and conclusions}

All the contributions to the Lamb shift considered in this paper are listed in Table~\ref{table}. Their sum 
produces the following theoretical prediction for the Lamb shift
\bea
\label{El2}
&&
\Delta E_L^{\rm this\,work}=
\left[
206.0243(30)
-5.2270(7)
\frac{ r_p^2 }{\mathrm{fm^2}}+0.0455(125)
\right]
\,\mathrm{meV}
\,.
\eea
The first ten terms in Table~\ref{table} are those associated to a pure QED-like computation assuming the proton to be point-like. Their sum is the first term in Eq.~(\ref{El2}), and 
its error is the estimate of the ${\mathcal O}(m_r\al^6)$ effects. The second term in Eq.~(\ref{El2}) encodes 
all the corrections proportional to the proton radius x)-xiii) entries in Table~\ref{table}). The error of the coefficient of the term proportional to $ r_p^2$ is the 
estimated size of the ${\mathcal O}(m_r\al^6 (m_{\mu}^2 r_p^2 ))$ terms. The last term encodes the $ r_p^2$-independent hadronic effects. The error is the assigned uncertainty due to unknown terms of ${\mathcal O}(m_r\alpha^5\frac{m_{\mu}^3}{m_{\rho}^3})$. We emphasize that a partial incorporation of subleading corrections in $\al$ to the above expression will not improve the precision of the result (unless there are 
arguments to argue that such contributions are the dominant ones), as the uncertainty is still dominated by unknown parametric terms of order $m_r\al^6$. For an account of some of these corrections see~\cite{Antognini:2013jkc}.

In order to obtain our primary result Eq. (\ref{El1}), the first and last term of Eq.~(\ref{El2}) has been added and the error combined in quadrature. With this, 
together with the experimental result in Eq.~(\ref{DeltaEexp}), we obtained in Ref.~\cite{Peset:2014yha} the value for the proton radius quoted in Eq.~(\ref{rpfinal}),
where the theoretical and experimental errors have been combined in quadrature. Nevertheless, the latter is completely subdominant with respect to the total error, which is fully dominated by the hadronic effects. In this 
respect it is also convenient to present our result in the following way
\begin{eqnarray}
\nn
&&\Delta E_L=
206.0243\,\mathrm{meV}\\
&&
-\left[\frac{1}{\pi}\frac{m_r^3\al^3}{8}\right]\frac{\al}{m_p^2}\frac{ r_p^2}{\mathrm{fm}^2}
\left[47.3525
+35.1491\al+47.3525\alpha^2\ln(1/\alpha)\right]\nonumber\\
&&
\nn+\left[\frac{1}{\pi}\frac{m_r^3\al^3}{8}\right]\frac{1}{m_p^2}
\left[
c_{3}^{\rm {had}}+
16\pi \alpha d_2^{\rm {had}}\right]
\\
&&+\mathcal{O}(m_r\alpha^6)
\,.
\label{El3}
\end{eqnarray}
Note that since $c_3^{\rm had} \sim \al^2$ and $\alpha d_2^{\rm {had}}\sim \al^2$, the third line of the previous 
equation encodes all the hadronic effects that are not related to the proton radius of order $\al^5$.  
This presentation of the result where $r_p$ and $c_3^{\rm had}$ are kept explicit could be important for the future.
In the long term (once the origin of the proton radius puzzle is clarified) the natural place where to get the proton radius is from the hydrogen Lamb shift and $c_3^{\rm had}$ (once the radius has been obtained) from the muonic hydrogen, since $c_3^{\rm had}$ is suppressed by an extra factor of the lepton mass. In this scenario a complete 
evaluation of the $\mathcal{O}(m_r\alpha^6)$ term may improve the precision of an eventual experimental determination of $c_3^{\rm had}$. Note that in this discussion we assume that we can determine $d_2^{\rm {had}}$ from alternative methods, like dispersion relations.

Finally, we profit this computation to give in the Appendix the exact $\al^5$ expression for the muonium spectrum, keeping the complete mass dependence, which can be easily deduced by changing $m_p \rightarrow m_{\mu}$ and $m_{\mu} \rightarrow m_{e}$, 
and setting the hadronic coefficients, $d_2^{\tau}$, and the electron vacuum polarization effects to zero.

\begin{table}[htb]
\addtolength{\arraycolsep}{0.15cm}
$$
\begin{array}{|c|c|c|c   |r l|}
 \hline 
{\rm i)}&{\mathcal O} (m_r \alpha^3)& V_{\rm VP}^{(0)} &  {\rm Eq.}\; (\ref{Evp}) & 205.&\hspace{-0.5cm}00737  
\\ \hline
{\rm ii)}&{\mathcal O} (m_r \alpha^4)& V_{\rm VP}^{(0)} & {\rm Eq.}\; (\ref{V02}) &1.&\hspace{-0.5cm}50795  
\\ \hline
{\rm iii)}&{\mathcal O} (m_r \alpha^4)& V_{\rm VP}^{(0)} & {\rm Eq.}\; (\ref{V0V0}) &0.&\hspace{-0.5cm}15090
\\ \hline
{\rm iv)}&{\mathcal O} (m_r \alpha^5)& V_{\rm VP}^{(0)} & {\rm Eq.}\; (\ref{V03})& 0.&\hspace{-0.5cm}00752
\\ \hline
{\rm v)}&{\mathcal O} (m_r \alpha^5)& V^{(0)}_{\rm LbL} & {\rm Eq.}\; (\ref{V0LbL})&   -0.&\hspace{-0.5cm}00089(2)
\\ \hline
{\rm vi)}&{\mathcal O} (m_r \alpha^4\times \frac{m^2_\mu}{m^2_p})& V^{(2,1)}+V^{(3,0)} & {\rm Eq.}\; (\ref{Ek})&  0.&\hspace{-0.5cm}05747
\\ \hline
{\rm vii)}&{\mathcal O} (m_r \alpha^5)& V^{(2,2)}_{\rm no-VP}+{\rm ultrasoft} & {\rm Eq.}\; (\ref{softUS}) &   -0.&\hspace{-0.5cm}71896
\\ \hline
{\rm viii)}&{\mathcal O} (m_r \alpha^5)& V^{(2,2)}_{\rm VP} + V^{(2,1)}\times  V^{(0,2)}_{\rm VP}+\cdots      &  {\rm Eq.}\; (\ref{V2VP})& 0.&\hspace{-0.5cm}01876
\\ \hline
{\rm ix)}&{\mathcal O} (m_r\alpha^6\times \ln (\frac{m_{\mu}}{m_{e}}))& V^{(2,3)}; c^{(\mu)}_D &
{\rm Eq.}\; (\ref{m6cDmutree})&  -0.&\hspace{-0.5cm}00127
\\ \hline 
{\rm x)}&{\mathcal O} (m_r\alpha^6\times \ln \al)& V^{(2,3)}_{\rm VP}; c^{(\mu)}_D& 
{\rm Eq.}\; (\ref{m6cDmuVP}) &  -0.&\hspace{-0.5cm}00454
\\[0.05cm] \hline \hline 
{\rm xi)}&{\mathcal O} (m_r \alpha^4\times m^2_r r^2_p)& V^{(2,1)}; c^{(p)}_D; r^2_p & {\rm Eq.}\; (\ref{Ek}) & -5.&\hspace{-0.5cm}19745
\frac{ r_p^2}{\rm fm^2}
\\[0.05cm] \hline
{\rm xii)}&{\mathcal O} (m_r \alpha^5\times m^2_r r^2_p)& V_{\rm VP}^{(2,2)}+\cdots; c^{(p)}_D; r^2_p & {\rm Eq.}\; 
(\ref{V2VP}) & -0.& \hspace{-0.5cm}02815
\frac{ r_p^2}{\rm fm^2}  
\\[0.05cm] \hline
{\rm xiii)}&{\mathcal O} (m_r \alpha^6\ln\al\times m^2_r r^2_p)& V^{(2,3)}; c^{(p)}_D; r^2_p
& {\rm Eq.}\; (\ref{r2al2})& -0.&\hspace{-0.5cm}00136
\frac{ r_p^2}{\rm fm^2}  
\\[0.05cm] \hline
{\rm xiv)}&{\mathcal O} (m_r \alpha^5\times \frac{ m_r^2}{m_{\rho}^2})& V_{\rm VP_{\rm had}}^{(2)}; d^{\rm had}_2 &{\rm Eq.}\; (\ref{VPhad}) &  0.&\hspace{-0.5cm}0111(2)
\\ \hline
{\rm xv)}&{\mathcal O} (m_r \alpha^5\times \frac{m^2_r}{m^2_{\rho}}\frac{m_{\mu}}{m_{\pi}})& V^{(2)}; c^{\rm had}_3 & {\rm Eq.}\; (\ref{ETPE}) &  0.&\hspace{-0.5cm}0344(125)
\\ \hline
\end{array}
$$
\caption{{\it The different contributions to the Lamb shift in muonic hydrogen in } meV {\it units.}}
\label{table}
\end{table}

\medskip

{\bf Acknowledgements} \\ 
This work was supported in part by  the Spanish grants FPA2013-43425-P, FPA2011-25948 and SO-2012-0234 and the Catalan grant
SGR2014-1450.

\newpage

\appendix
\section{Muonium spectrum}

We profit from the results obtained in this work to give the spectrum for the muonium bound state ($\mu e$) for general quantum numbers at ${\mathcal O}(m_r\al^5)$. We first exchange the proton by the muon and the muon by the electron. Then, the main difference with muonic hydrogen is the lack of hadronic contributions, as well as the fact that all electron vacuum polarization effects can be eliminated, in particular this implies that 
the static potential becomes trivial. Thus, we are only left with the relativistic corrections to the potential which come from Eqs. (\ref{VbcD}) and (\ref{tildeV1loop}) plus the energy coming from the kinetic term and the ultrasoft effect. 
The ultrasoft correction to the energy only depends on the reduced mass, and so it will be the same as the one for the muonic hydrogen in Eq. (\ref{Eus1}).
Altogether, for a given energy level we get
\bea
E_{nljj_e}&=&-\frac{m_r\al^2}{2n^2}+(\delta E^{V^{(2,1)}}_{nljj_{e}}+\delta E^{V^{(3,0)}}_{nl})+(\delta E_{nljj_{e}}^{V^{(2,2)}_{\rm no-VP}}+\delta E^{\rm US}_{nl})\nn\\
&=&-\frac{m_r\al^2 }{2n^2}
\nn
\\
&&
+\frac{m_r\alpha^4 }{n^3}\left[\frac{m_r^2}{2m_e^2}\left\{\delta_{l0}+\frac{3}{4n}-\frac{2}{2l+1}+\frac{(1-\delta_{l0})}{l(l+1)(2l+1)}d_{j_e,l}+2\frac{m_e}{m_\mu}\left(\frac{5}{8n}-\frac{2+\delta_{l0}}{2l+1}\right.\right.\right.\nonumber\\
&&+\left.\left.\left.\frac{8}{3}\delta_{l0}\delta_{s1}+\frac{(1-\delta_{l0})\delta_{s1}}{2l(l+1)(2l+1)}(c_{j,l}+4h_{j,l})\right)\right.\right.\nonumber\\
&&+\left.\left.\frac{m_e^2}{m_\mu^2}\left(\delta_{l0}+\frac{3}{4n}-\frac{2}{2l+1}+\frac{(1-\delta_{l0})}{l(l+1)(2l+1)}\left(2\delta_{s1}h_{j,l}-d_{j_e,l}\right)\right)\right\}\right.\nn\\
&&+\left.\frac{\alpha}{\pi}\left\{
\delta_{l,0}\left(-\frac{4}{3}\left(\ln R(n,l)+2\ln\alpha\right)+\frac{10}{9}\right)-(1-\delta_{l,0})\frac{4}{3}\ln R(n,l)
\right.\right.\nn\\
&&+\left.\left.\frac{m_r^2}{2m_e^2}\left\{\frac{4}{3}\left(-\frac{2}{5}+\ln\left(\frac{m_e^2}{m_r^2}\right)\right)\delta_{l,0}+\frac{1-\delta_{l,0}}{l(l+1)(2l+1)}d_{j_e,l}\right.\right.\right.\nn\\
&&+\left.\left.\left.\frac{m_e^2}{m_\mu^2}\left(\frac{4}{3}\left(-\frac{2}{5}+\ln\left(\frac{m_\mu^2}{m_r^2}\right)\right)\delta_{l,0}+\frac{1-\delta_{l,0}}{l(l+1)(2l+1)}(2\delta_{s,1}h_{j,l}-d_{j_e,l})\right)\right.\right.\right.\nn\\
&&+\left.\left.\left.\frac{m_e}{m_\mu}\left(-\frac{10}{3}\delta_{l,0}-\frac{14}{3}\frac{1-\delta_{l,0}}{l(l+1)(2l+1)}+\frac{14}{3}\delta_{l,0}\left(1-\frac{1}{n}+2k(n)+2\ln\left(\frac{2\alpha }{n}\right)\right)\right.\right.\right.\right.\nn\\
&&+\left.\left.\left.\left.\frac{16}{3}\delta_{s,1}\delta_{l,0}+\frac{1-\delta_{l,0}}{l(l+1)(2l+1)}(c_{j,l}+2h_{j,l})\right)+\frac{2m_e^2\delta_{l,0}}{m_e^2-m_\mu^2}\left(\frac{m_\mu}{m_e}\left(\frac{1}{3}+\ln\left(\frac{m_e^2}{m_r^2}\right)\right)\right.\right.\right.\right.\nn\\
&&+\left.\left.\left.\left.\frac{m_e}{m_\mu}\left(\frac{1}{3}+\ln\left(\frac{m_\mu^2}{m_r^2}\right)\right)+\ln\left(\frac{m_e^2}{m_\mu^2}\right)(3-4\delta_{s,1})\right)\right\}\right\}\right],
\eea
where $c_{j,l}$, $h_{j,l}$ and $d_{j_e,l}$ have been defined in Eqs. (\ref{c})-(\ref{d}), and the first and second parenthesis in the right hand side of the first equality contain the ${\mathcal O}(m_r\alpha^4)$ and ${\mathcal O}(m_r\alpha^5)$ contributions respectively. Note that the exact mass dependence has been kept in this expression to order $\al^5$. 

The expressions for the potential of muonium can also be found in Ref.~\cite{GRS}. One could be worried that the potential is different to the one we use. 
The reason for this difference is that they obtain the potential by matching on-shell S-matrix elements (and by a change in the renormalization scheme of the ultrasoft computation), still their potential is equivalent to ours through field redefinitions, and yields the same physical results. In particular, for spin-independent states the result for the energy shift can already be found in Eqs. (2.12) and (2.13) of 
that reference.

\end{document}